\documentclass{aa}

\usepackage{graphicx}
\usepackage{txfonts}
\usepackage{xcolor}
\usepackage{xspace}
\usepackage{hyperref}

\newcommand{\betapic}{$\beta$ Pictoris\xspace}

\begin{document}

   \title{Exocomets of $\beta$ Pictoris I:}
   \subtitle{Exocomet destruction, sodium and disk line variability in 17 years of HARPS observations}

   \author{H.J. Hoeijmakers
          \inst{1}
          \and K.P. Jaworska
          \inst{1}
          \and B. Prinoth
          \inst{1}
          }

    \institute{
    Lund Observatory, Division of Astrophysics, Department of Physics, Lund University, Box 118, 221 00 Lund, Sweden
             }

   \date{}

  \abstract
   {The young \betapic system has been monitored with high-resolution optical spectrographs for decades. These observations have revealed strongly variable, stochastic absorption in the Ca\,II H\&K lines attributed to in-falling cometary bodies.}
   {Since 2003, over 9000 HARPS observations of \betapic have been taken and many of these have not yet been used for exocomet studies. We aim to search these spectra for new exocomet phenomenology enabled by the long time coverage and large volume of this dataset.}
   {We systematically carry out telluric correction of the HARPS spectra using {\tt molecfit}, compare multi-year observations at the wavelengths of the Ca\,II and Na\,I lines and use a Bayesian fitting algorithm to extract exocomet line parameters. We explore the usage of an unbiased reference spectrum with which to calibrate the continuum, and investigate Keplerian orbital solutions to observed exocomet acceleration.}
   {We find a general absence of exocometary sodium line absorption, with only two instances of clear ($\sim 2\%$ deep) exocometary sodium out of 198 nights of observation, as well as a weaker ($\sim 1\%$) feature that persists over 13 nights in 2004. We find that these events occur during times of exceptionally deep Ca II absorption, at the same red-shift, implying that strongly Ca\,II-evaporating exocomets also exhibit detectable levels of Na\,I, in spite of the vast majority of Na\,I being rapidly photo-ionised in close proximity to the star. We find long-lived Ca\,II absorption in 2017 and 2018 that persists on a timescale of a year, which may be difficult to explain with the classical exocomet model. Finally, we investigate two strongly accelerating, blue-shifted exocomet features observed in 2019 that show strong and sudden departures from Keplerian motion, suggesting rapid changes to the dynamics of the exocomet cloud. We hypothesize that this is caused by the destruction of the comet nuclei shortly after their periastron passages.}
   {}


   \maketitle
%

\section{Introduction}
The young, bright and nearby \betapic system presents an iconic planet formation laboratory. Two massive gas giants \citep{Lagrange2009,Lagrange2019} on intermediately close orbits of 2.7 and 9.9 AU \citep{Lacour2021} create a dynamically multi-faceted system, with a structured outer debris disk. The disk was immediately hypothesized to be associated with planet formation upon its discovery \citep[][]{Aumann1985,Smith1984}, long before the eventual detections of \betapic b and c, and has been a target of intense study ever since. Observations of the infrared excess emission of the disk have shown that there exist multiple dust components, both far away from the star as well as within the inner few AU \citep{Okamoto2004,Chen2007,Lu2022}. Recent observations by JWST/MIRI MRS have shown that the dust emission is variable over decade-timescales, forming evidence for large planetesimal collisions, including close to the star \citep{Chen2024}. The star is also surrounded by gas-phase atoms and ions at various distances, visible in numerous absorption lines of calcium, sodium, iron and other metals \citep[e.g.][]{VM1986,Hobbs1988}.\\

\subsection{Exocomets}
\noindent An emblematic property of \betapic is the existence of planetesimals on highly eccentric orbits that violently sublimate on close approach to the star. These objects, classically referred to as Falling Evaporating Bodies (FEBs) but now commonly referred to as exocomets, were first observed in the mid 1980's owing to their strong line-absorption near the wavelengths of the Ca\,II H\&K lines \citep[e.g.][]{Ferlet1987} as well as other metal ion lines \citep[e.g.][]{Lagrange1987}, that vary on short time scales and occur at large Doppler shifts. Throughout the 1990's many targeted spectroscopic observations and theoretical studies were carried out to further develop and cement the exocomet model \citep[see][for a review]{Beust2024}. Final confirmation of the exocomet hypothesis has recently come in the form of detection of exocomet transits in broad-band photometry from the Transiting Exoplanet Survey Satellite (TESS) \citep{Zieba2019,Pavlenko2022,Lecavelier2022}. These transit light-curves have distinct shapes consistent with material arranged in a cometary tail, as originally predicted by \citet{Lecavelier1999}.\\

\noindent The initial detections of massive close-in exoplanets using high-resolution spectroscopic monitoring of bright stars \citep{Mayor1995} spurred the development of dedicated, highly stabilized high-resolution spectrographs. Since its inception, the HARPS high-resolution spectrograph on ESO's 3.6-m telescope in La Silla observatory \citep{Pepe2002} has been used to systematically observe the \betapic system, both for the purpose of exocomet monitoring as well as later establishing the masses of \betapic b and c \citep[e.g.][]{Lagrange2012}. By 2011, over 1000 HARPS spectra of \betapic had been obtained, which were analyzed statistically for the first time by \citet{Kiefer2014}. Assuming the obscuring cloud of a comet to be homogeneous, fitting for the line ratio of the Ca\,II H\&K lines allows the optical depth as well as the fraction of the stellar disk that is covered by the cloud to be constrained \citep[e.g.][]{Lagrange1989}. \citet{Kiefer2014} found that there exist two populations of exocomet absorption features: Events with a broad distribution in radial velocities, relatively wide absorption lines and small fill fractions, and lower red-shift events with narrower absorption lines and larger fill fractions. This was interpreted as evidence that these populations tend to transit at different distances from the star, as larger clouds can be sustained at larger distances from the star where radial velocities are typically smaller. \citet{Kiefer2014} also investigated the region of the neutral sodium (Na\,I) doublet near 589 nm, and confirmed the existence of a static absorption line attributed to the circumstellar disk \citep{VM1986}.\\

\bigskip

\noindent This paper is the first in a series in which we investigate exocomets in the $\beta$ Pictoris system. Here, we introduce a new analysis methodology of the HARPS spectra and highlight a number of general aspects of the observed exocomets that we consider novel. We also focus on two accelerating exocomet features in 2019 and apparent exocomet activity in the 589\,nm sodium doublet, while leaving a comprehensive statistical analysis of the large sample of observed Ca\,II features \citep[similar to e.g.][]{Kiefer2014} for a later study.

\section{Methods}
\subsection{Data}
The \betapic system has been observed extensively since 2003 with the HARPS high-resolution spectrograph on ESO's 3.6 m telescope in La Silla observatory. The public data archive lists 9133 spectra obtained until April 2020. We download these spectra\footnote{This study is based on data obtained from the ESO Science Archive Facility with DOI: \url{https://doi.eso.org/10.18727/archive/33}.}, rejecting a small number of spectra that are incorrectly associated with \betapic or have excessive noise, leaving a total of 9071 spectra taken as part of 26 programs on 198 unique nights, with a total exposure time of 159.4 hours. At present, we do not aim to carry out a comprehensive statistical analysis, but we provide annual statistics of these observations in Table \ref{tab:obs_yearly}.\\

\begin{table}
  \caption[]{Statistics of archival HARPS observations grouped by year.}
    \label{tab:obs_yearly}
    \centering
     \begin{tabular}{p{0.15\linewidth}ccc}
        \hline
        \hline
        \noalign{\smallskip}
        Year & $N$ nights & $N$ spectra & $t_{\textrm{exp}}$ (h) \\
        \noalign{\smallskip}
        \hline
        \noalign{\smallskip}

2003 & 24 & 137 & 4.8 \\
2004 & 29 & 116 & 4.0 \\
2005 & 7 & 28 & 0.2 \\
2006 & 5 & 10 & 0.2 \\
2007 & 8 & 30 & 0.4 \\
2008 & 11 & 355 & 6.4 \\
2009 & 10 & 349 & 7.0 \\
2010 & 1 & 60 & 1.5 \\
2011 & 4 & 207 & 4.3 \\
2012 & - & - & - \\
2013 & 18 & 354 & 9.5 \\
2014 & 8 & 404 & 5.9 \\
2015 & 6 & 160 & 1.6 \\
2016 & 7 & 420 & 3.8 \\
2017 & 20 & 2891 & 53.6 \\
2018 & 23 & 1380 & 24.1 \\
2019 & 9 & 1175 & 21.6 \\
2020 & 8 & 995 & 10.6 \\
\hline
\textbf{Total} & \textbf{198} & \textbf{9071} & \textbf{159.4} \\ 
     \end{tabular}

\end{table}

\noindent Previous studies \citep[e.g.][]{Kiefer2014} have used the one-dimensional, blaze-corrected and stitched 1D spectra produced by the HARPS Data Reduction Software (DRS). In this study, we follow our previous practice \citep[e.g.][]{Hoeijmakers2020,Prinoth2022,Hoeijmakers2024} by using the individual extracted spectral orders (e2ds files) instead, to avoid unnecessary re-interpolation that the pipeline carries out to produce data on a uniform wavelength grid and to average wavelength regions where orders overlap. Using the individual spectral orders is operationally nearly identical to using the 1D spectra, with the caveat that some wavelength ranges are covered by two orders (this is the case for the Ca\,II H\&K lines), and the fact that the stellar continuum is modified by the blaze function - which does not affect our methodology (see Methods). As a result, we are able to directly use the count values in each of our spectral channels as representative of the flux that was recorded by the detector. Assuming that our spectra are in the photon-noise limit because \betapic is a very bright (V=3.9) star, the uncertainties are estimated as the square root of the recorded count levels -- mitigating the fact that the reduced 1D spectra are not provided with estimates of the uncertainty at each wavelength.\\

\noindent Using the individually extracted orders (e2ds files) requires the application of the barycentric velocity correction for each spectrum in the dataset. For this, we use the value of the barycentric velocity provided by the pipeline. This leads to a unique wavelength axis for each spectrum in the dataset, even if the wavelength solution was constant in each night of data. Calculation of time-average spectra (e.g. the reference spectrum, see section \ref{sec:reference_spectrum}) still requires an interpolation onto a common wavelength grid, but for all other purposes in this paper we retain a separate wavelength axis for each order of each of the spectra in the dataset.\\

Note that for comparing spectra obtained at different times (e.g. to compute a reference spectrum, see section \ref{sec:reference_spectrum}) we need to account possible variations in the broadband continuum, e.g. due to atmospheric dispersion or flat-field changes. As \citet{Kiefer2014} makes use of pipeline calibrated spectra, such variations may have been largely removed, but the e2ds spectra can attain end-to-end variations in the flux level of the continuum by 5\% to 10\%, that we correct following a methodology similar to our previous work \citep[e.g.]{Hoeijmakers2020}, by fitting fourth order polynomials to the residuals of the time-average spectrum using a Least Absolute Deviation (LAD) optimization, while iteratively rejecting outliers and bad regions (telluric residuals, varying exocomet absorption lines).

\subsection{Telluric correction}
\begin{figure}
    \centering
    \includegraphics[width=0.5\textwidth,trim={1.5cm 1cm 1cm 2cm},clip]{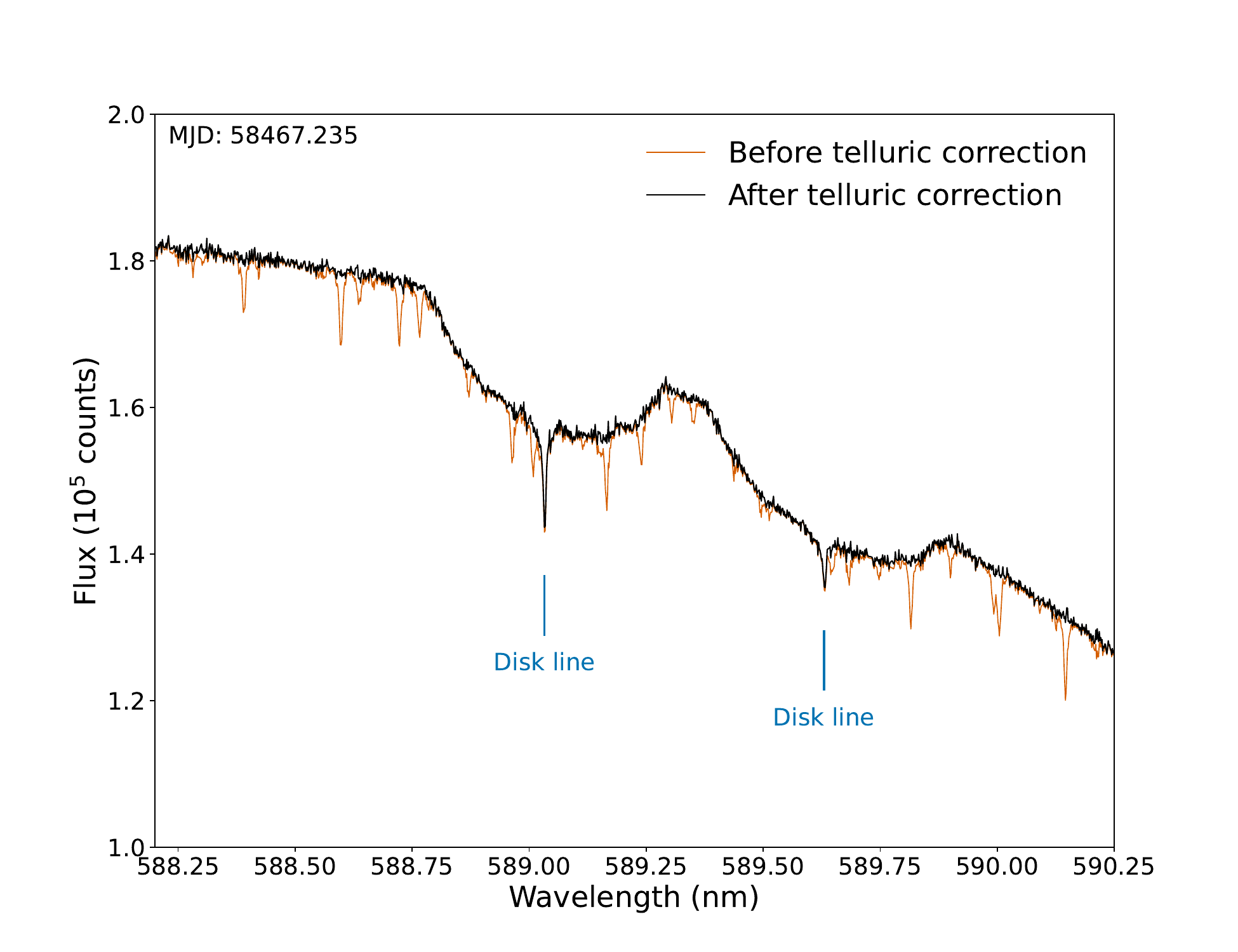}
    \caption{Example of telluric correction near the neutral sodium doublet. The doublet from the circumstellar disk is readily visible.}
    \label{fig:telcor}
\end{figure}

Besides Ca\,II, we intend to search for exocomet activity in the Na\,I doublet near 589 nm. This region is contaminated by telluric water lines, hindering the identification of exocomet sodium lines. We remove telluric contamination systematically from all spectra in the dataset using {\tt molecfit} \citep{Smette2015,Kausch2015}. Because the line-spread function of HARPS varies from the short to the long wavelength side of each order \citep[e.g.][]{Zhao2021}, and because we are interested only in correcting the water lines at the wavelengths of the sodium doublet, we restrict {\tt molecfit} to 10 water lines immediately adjacent to Na\,I lines. We avoid all water lines inside the broad photospheric Na\,I lines because strong exocomet sodium lines could bias telluric fitting. A typical correction is shown in Figure \ref{fig:telcor}.

\subsection{Determining a reference spectrum}\label{sec:reference_spectrum}
Similar to \citet{Kiefer2014}, we proceed to use the fact that exocomet absorption is highly variable in strength and wavelength to identify a reference spectrum of the system. The principle purpose of this is to correct for the absorption lines of the circumstellar gas -- which are assumed to be static but partially blend with exocomet features, and to reliably identify exocomet absorption in the steep wings of the stellar absorption lines. To do this, we follow a similar approach to \citet{Kiefer2014}, by estimating the mean flux at each wavelength channel using only their highest flux values -- rejecting flux values that are potentially affected by stochastic exocometary absorption. The method adopted by \citet{Kiefer2014} assumes that all flux values in a spectral channel are drawn from the same random distribution. However the large number of spectra available today span a wide range of signal-to-noise ratios (from a few dozen to over 200 in the center of the order covering Ca\,II K. We therefore adopted the following method that allows for the variable uncertainties between spectra:\\

Consider that the data consists of 9071 exposures $0 .. i .. 9071$. Assume that in the absence of exocomet absorption, the flux $X_{ij}$ (of exposure $i$ in each wavelength channel $j$) is drawn from a normal distribution centered on a mean flux $\mu_j$ with a standard deviation $\sigma_{ij}$, equal to the measurement uncertainty (photon noise) of that particular exposure. Taking the mean of this collection of samples $\langle X \rangle$ would normally yield an accurate estimate for $\mu_j$. However, due to the stochastic presence of exocomet absorption, an unknown number of these flux values will be depressed, biasing the estimate of $\mu_j$ towards lower values. Define a threshold $T_j$ above which the flux samples in the wavelength channel are unlikely to be affected by exocomet absorption, and thus drawn from the unbiased distribution. This collection of samples forms a truncated normal distribution, and the true mean $\mu_j$ can be obtained from the mean of the truncated sample \citep{Tallis1961} as follows:

\begin{equation}\label{eq:debias}
    \langle X_{ij}|X>T \rangle = \mu_j + \sigma_{ij} \frac{\phi(\alpha)}{1- \Phi(\alpha)},
\end{equation}

where $\phi$ and $\Phi$ respectively are the probability density function (PDF) and cumulative density function (CDF) of the normal distribution, and $\alpha$ is the standardized truncation threshold:

\begin{equation}
    \alpha = \alpha_{ij} = \frac{T-\mu_j}{\sigma_{ij}},
\end{equation}

i.e. the number of standard deviations that the truncation threshold is away from the mean. But as $\mu_j$ is unknown in the presence of exocomets, equation \ref{eq:debias} is to be solved iteratively: For each sample $i,j$, assume an estimate for $\mu_j$ to compute $\alpha$. Subtract the term $\sigma_{ij}\frac{\phi}{1-\Phi}$ from each element in the truncated sample $X_{ij} > T$, and take the mean to obtain an estimate of $\mu_j$, and repeat until convergence.\\

At this stage, the computation of the reference spectrum still depends on the choice of threshold value $T$, which we take to be the 80$^\mathrm{th}$ percentile of the entire distribution of flux values $X_{ij}$. To arrive at this choice, we varied $T$ between between the 10$^\mathrm{th}$ to 98$^\mathrm{th}$ percentile, and determined that the reference spectrum obtained by truncating the samples below 80$^\mathrm{th}$ percentile lies within 2\% of the reference spectrum that would be obtained when using the 98$^\mathrm{th}$ percentile, at all wavelengths beyond 15 km/s from the center of the disk line, this providing an accurate estimate of the reference spectrum based on a statistically large ($\sim 2000$) number of samples (see Fig. \ref{fig:percentiles}.

\begin{figure}
    \centering
    \includegraphics[width=0.5\textwidth,trim={1.2cm 0cm 1cm 0cm},clip]{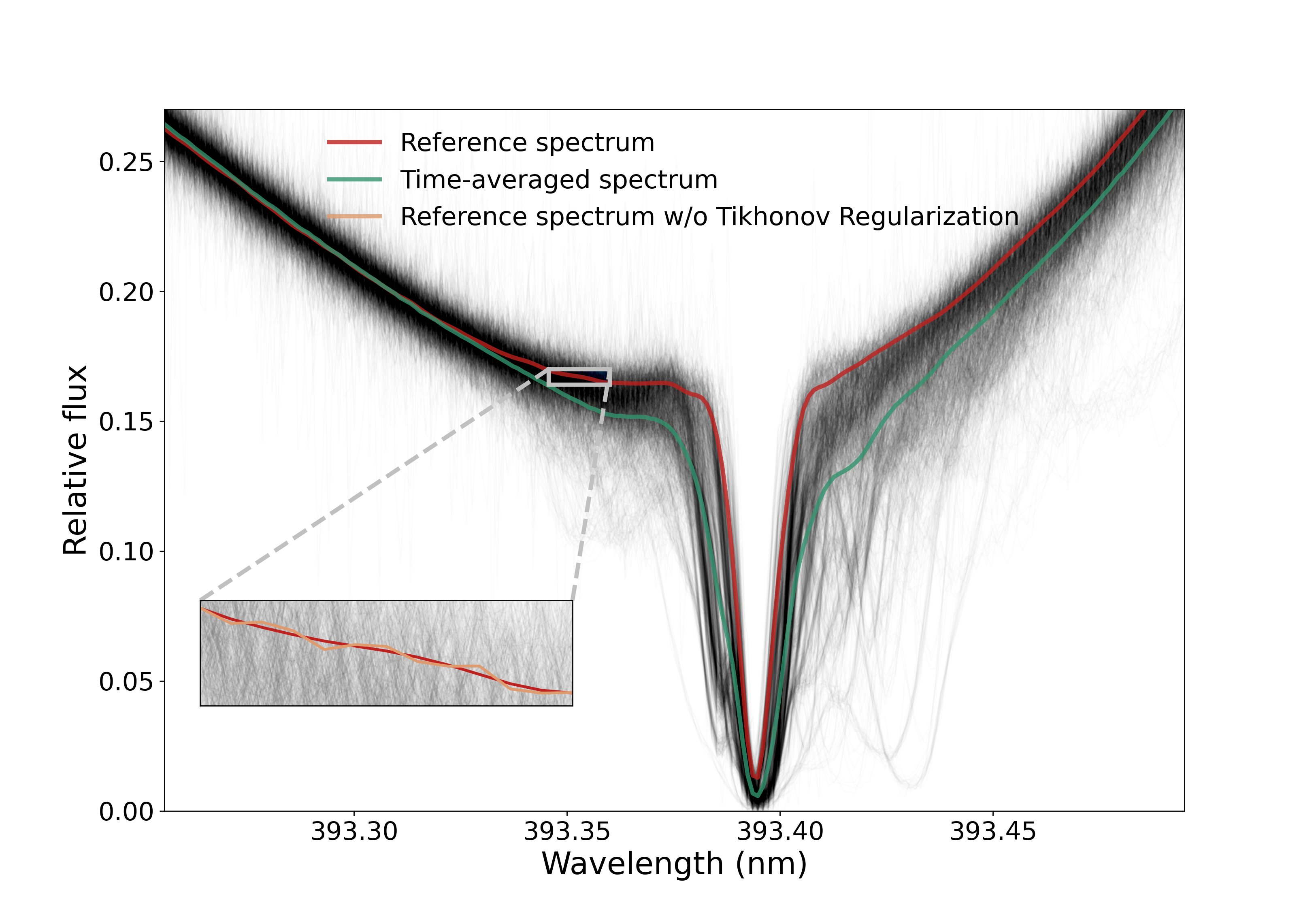}
    \caption{Reference spectrum obtained following the methodology described in section \ref{sec:reference_spectrum} (red), compared to the simple time averaged spectrum (green) and the HARPS data (black). Smoothing using Tikhonov regularization reduces the noise in the reference spectrum at the 0.5\% level (inset). As expected for an unbiased estimator, the reference spectrum tends to the time-averaged spectrum at wavelengths where no exocomet activity is present.}
    \label{fig:F_ref}
\end{figure}

However, like the reference spectrum obtained by \citet{Kiefer2014}, the estimated reference spectrum is affected by noise. Because of this, we restrict the above formalism only to the 50\% of spectra that have a higher than average signal to noise ratio (in the range of 100 to 200 measured in the center of order 6). In addition, we use Tikhonov regularization\footnote{See e.g. \citep{Murata2022} for an application of Tikhonov regularization in astrophysical image analysis.} to obtain a weakly smoothed approximation to the noisy reference spectrum by solving the system:

 \begin{equation}
     \mu = \left(I-\lambda D^T D\right)\mu_{\mathrm{smooth}},
 \end{equation}

where $I$ is the identity matrix and $\lambda D^T D$ is a regularization term, in which $D$ is a sparse matrix populated by the finite difference approximation of the fourth order derivative (1,-4,6,-4,1) on the diagonal and $\lambda$ set to unity. The regularization term effectively penalizes roughness in the fourth derivative of the reference spectrum, acting to smooth variations in adjacent wavelength channels while preserving narrow spectral lines. We determined that the smoothed reference spectrum is within 0.5\% of the un-smoothed reference spectrum at all wavelengths greater than 7 km/s away from the center of the disk lines. The reference spectrum thus obtained is shown in Fig. \ref{fig:F_ref}.

\subsection{Exocomet line fitting}
We follow \citet{Kiefer2014} in constructing an analytical model of the exocomet absorption lines. The absorbing cloud is approximated as a homogeneous medium with an optical depth at the line center of $\tau_0$. The line shape is assumed to be Gaussian, making the relative absorption equal to:

\begin{equation}
    A = 1 - e^{-\tau_0 \Phi },
\end{equation}

where $\Phi$ is the Gaussian line profile: 

\begin{equation}
    \Phi = e^{-\frac{(\lambda-\lambda_0)^2}{2 \sigma^2}},
\end{equation}

with $\lambda_0$ equal to the center wavelength of the line, and $\sigma$ the line width.\\

\begin{figure}
    \centering
    \includegraphics[width=0.5\textwidth,trim={2cm 2cm 0cm 0cm},clip]{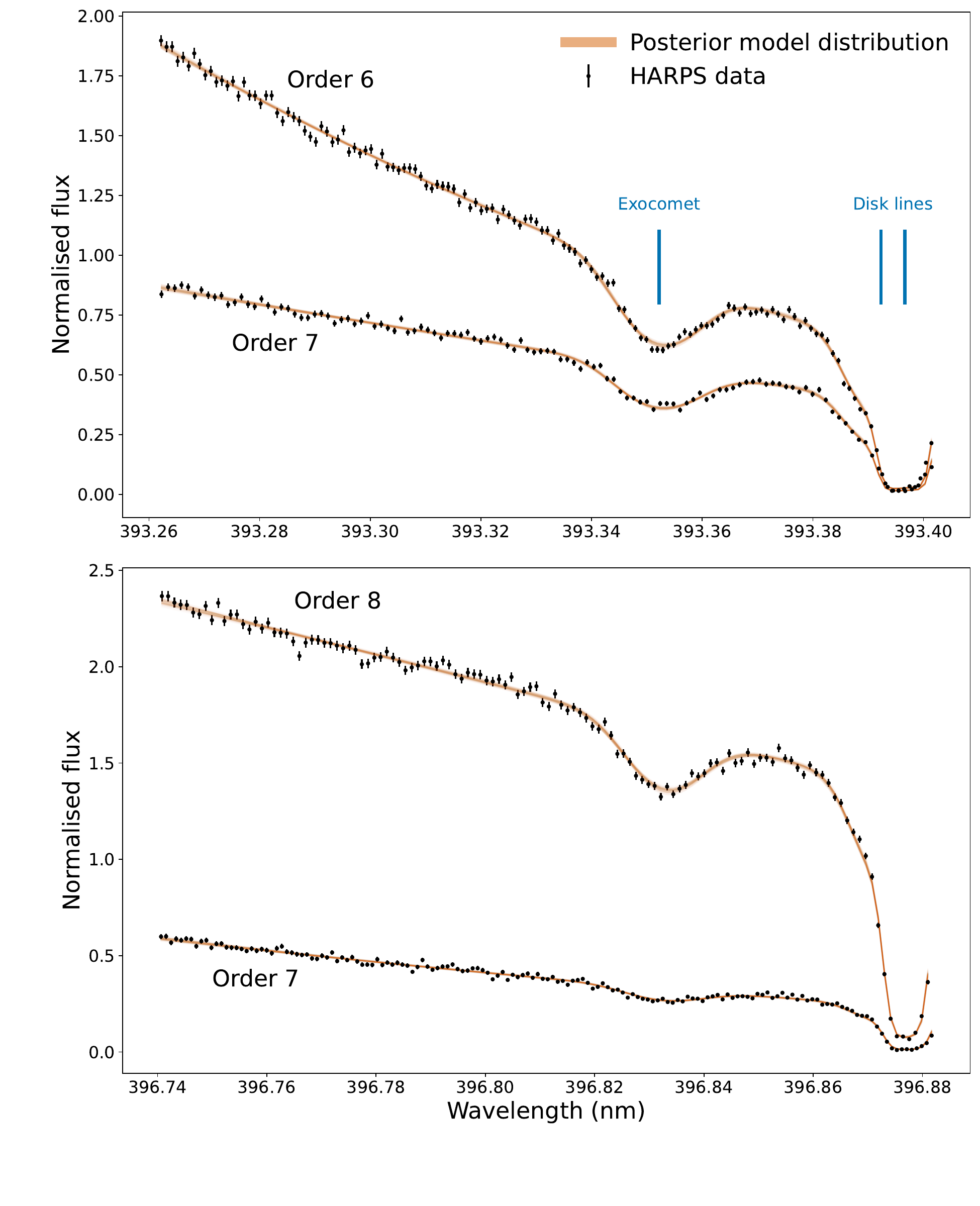}
    \caption{Example of a fit to the Ca\,II doublet (K-line in the top panel and H-line in the bottom panel) of a blue-shifted exocomet observed in the night of December 10, 2019. The exocomet is modeled as a single component, and the disk line as a blend of two. The narrow fitting region is restricted to the blue side of the disk line to avoid additional unrelated exocomet components. Because the Ca\,II lines are covered by three spectral orders of HARPS, the fit is carried out over four wavelength regions simultaneously. The posterior distribution of models is shown by the orange lines. Posterior distributions for model parameters are shown in Fig. \ref{fig:cornerplot}.}
    \label{fig:fitting_example}
\end{figure}

\begin{figure}
    \centering
    \includegraphics[width=0.5\textwidth,trim={0cm 0cm 0cm 0cm},clip]{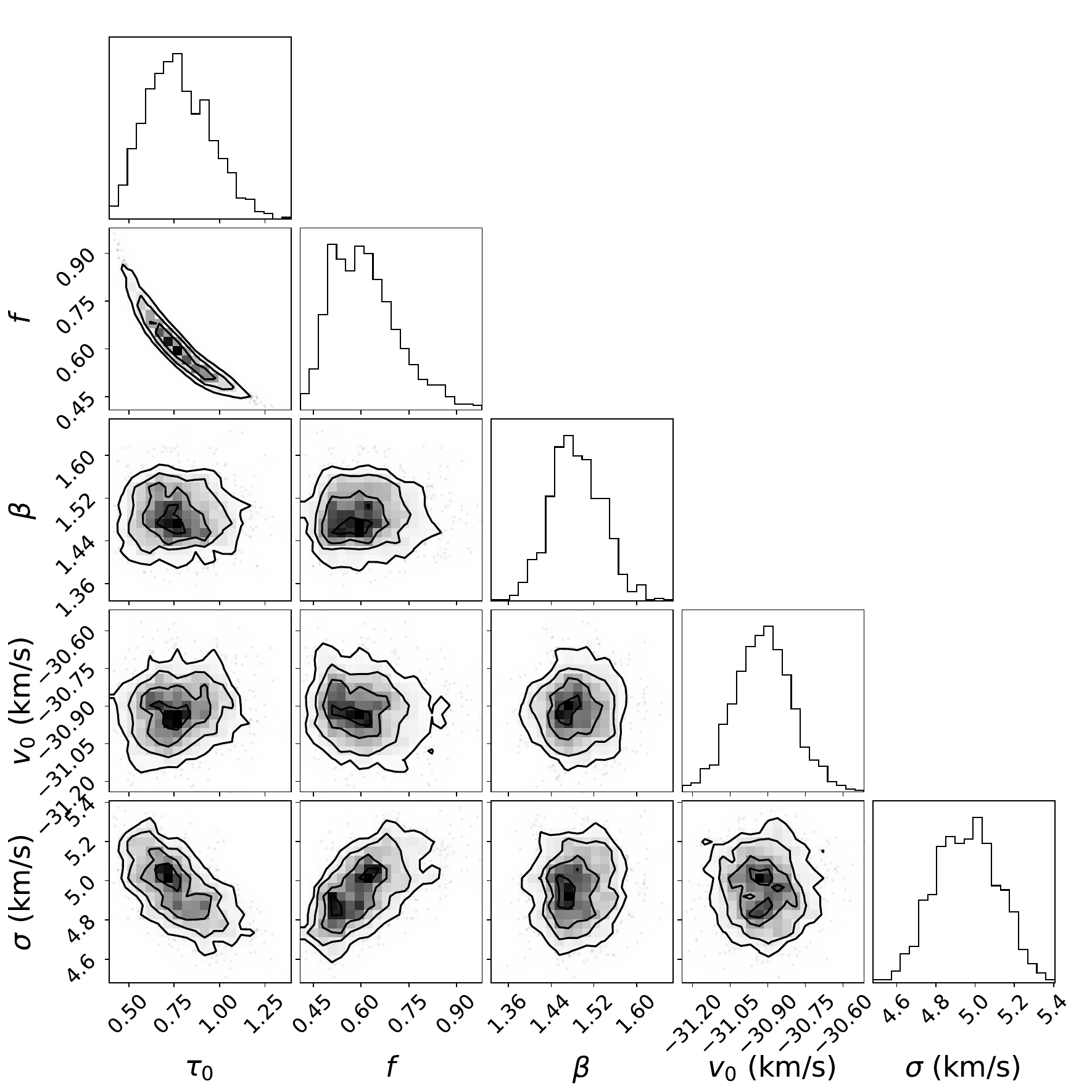}
    \caption{Posterior distributions for five relevant model parameters of the fit shown in Fig. \ref{fig:fitting_example}, out of a total of 27 free parameters (see Methods). Note that typical for fits of exocomet features that are not optically thick, there is a degeneracy between the peak optical depth $\tau_0$ and the fill factor $f$. However in this example, $f$ is sufficiently tightly constrained to conclude that the star is not fully covered by the cloud. Also note that the radial velocity of the cloud is constrained to within an uncertainty of about 100 m/s, and that the $1 \sigma$ line width is approximately 5 km/s.}
    \label{fig:cornerplot}
\end{figure}

\noindent For both calcium and sodium, we are studying line doublets with ratios in their oscillator strengths equal to 2. For small optical depth $\tau_0 < 1$, the line ratio therefore approaches 2. For large optical depths $\tau_0 \gg 1$, both lines saturate, resulting in equal line depths. However, if the cloud does not fully cover the stellar disk, the absorption lines do not extinguish the entirety of the stellar flux at the line center despite the large optical depth. This means that observations of line doublets are sensitive to the fill factor $f$ of the stellar disk -- a proxy for the radial extent of the cloud if assuming point-symmetry around the line of sight. We thus model the fraction of transmitted flux in one line $i$ as follows, equivalent to \citet{Kiefer2014} and more recently \citet{Vrignaud2024} using a large collection of Fe\,II lines simultaneously:
\begin{equation}
F_i(\lambda) = 1 - f\left( 1 - e^{-\tau_0 \Phi } \right). \label{eq:linemodel}
\end{equation}

\noindent The fractional line depth measured at the center of the line is $D = f\left( 1 - e^{-\tau_0} \right)$.\\

\noindent Because we are using non-normalized spectra, the continuum flux $F_c(\lambda)$ is arbitrarily scaled and slowly varying with wavelength because our target lines are situated in the cores of the broad stellar lines, and the spectra extracted by the pipeline are not flux-calibrated. We therefore assume that the continuum is described by a low-order polynomial function $F_c = c_0 + c_1 (\lambda-\lambda_0) + ... + c_n (\lambda-\lambda_0)^n$, where we typically set $n=2$ or $n=3$ to get a satisfactory fit without a dramatic increase in the number of free parameters. As there may be multiple (overlapping) exocomet features in any spectrum, we allow multiple components to be fit simultaneously:

\begin{equation}
    F(\lambda) = F_0(\lambda)F_1(\lambda)...F_N(\lambda)F_c(\lambda). \label{eq:fullmodel}
\end{equation}

\noindent The doublet lines we are targeting are widely separated in wavelength compared to the typical line widths and velocities involved. We wish to restrict the range of wavelengths over which we fit Eq.\,\ref{eq:fullmodel} for the continuum polynomial approximation to hold. Moreover, the Ca\,II H\&K lines are covered by multiple spectral orders of HARPS simultaneously, with the K-line appearing in orders 6 and 7, and the H-line in orders 7 and 8. This means that for fitting the Ca\,II doublet, we include four narrow wavelength slices, each with its own continuum. The H line has a line strength that is half of that of the K line, so we fix its optical depth accordingly. The same situation applies to the Na doublet, but these lines are covered by a single spectral order, meaning that only two wavelength slices need to be included in a fit.\\

\noindent For any one spectrum, the model therefore has the following free parameters when fitting $N$ line components:
\begin{itemize}
\item $N$ center wavelengths $\lambda_0$ of each line component.
\item $N$ line widths $\sigma$.
\item $N$ center optical depths $\tau_0$ of one of the doublet lines (the other being fixed via the known ratio of oscillator strengths).
\item $N$ surface fill fractions $f$.
\item $n+1$ polynomial components for each of the slices. For fitting the Ca\,II lines, we have four slices, giving 12 to 16 free parameters.
\item A factor $\beta$ that uniformly scales the uncertainties to capture unmodeled variance (e.g. small exocomet features that were not fitted explicitly with a separate line component or telluric residuals). Using the same notation as in Section \ref{sec:reference_spectrum}, this effectively modifies the log-likelihood of spectrum $j$ as: $\log L_j \propto -\frac{1}{2} \Sigma_i \left( \frac{X_{ij} - F_{ij}}{\beta \sigma_{ij}} \right)^2$. This use of $\beta$ to scale uncertainties is common in exoplanet model fitting applications, including the fitting of transit light-curves that are prone to correlated noise \citep[e.g.][]{Winn2008}. A commonly used alternative method to capture unmodeled variance is by adding a constant term to the reported uncertainties, often referred to as a red noise or jitter term \citep[see][for an early application]{Pont2006}.
\item Optionally a wavelength shift $d\lambda$ between one slice and the others (i.e. the wavelength shift between the two lines of the doublet or a shift between the wavelength solutions of overlapping orders) to allow for any errors in the wavelength solution of the spectrograph. For the calcium lines this gives two extra free parameters, for the sodium lines one.
\end{itemize}

\noindent Because this model is analytical, we implement the model as an auto-differentiable function in \texttt{Jax}, and use NumPyro to sample from the prior distributions (see below) with a no U-turn sampler \citep{Betancourt2017,Bradbury2018,Bingham2019,phan2019composable}. An example of a best-fit model to a single exocomet feature with two components to describe the circumstellar disk line is shown in Fig. \ref{fig:fitting_example}. This is a model with 27 free parameters. \\

\noindent A key choice for any spectrum is the wavelength range to fit over, the number of line components to include and the definition of prior parameter distributions. There are nearly 10,000 spectra that we may wish to fit in this way, which would be prohibitively labor intensive. Using a relatively small subset of the present data, \citet{Kiefer2014} used an automated algorithm that chooses the number of exocomet components by evaluating and comparing the Bayesian information criterion. Our model would allow for a similar workflow, but in the present study we target particular exocomet signatures in two nights of observation in 2019, and the sodium lines which, as we will see, typically have only one coherent absorption signature shifted away from the disk line (see Results). Secondly, it is worth noting that exocomet features typically do not vary strongly between the spectra taken in any one night of observation. In theory, this means that the definition of components and priors can in most cases be done on a night-by-night basis. The computational efficiency offered by \texttt{Jax} makes it feasible to execute model fitting on large numbers of spectra, and we aim to use this framework to carry out more rigorous statistical analysis of exocomet absorption signatures in these data in a future study.\\

\noindent A final choice is the assumption on the systemic velocity, which sets the zero-point of any measurements of radial velocities of exocometary spectral lines or disk lines. An independent measurement of the radial velocity of the \betapic host star is complicated by its high rotation velocity and $\delta$ Scuti pulsations that cause time-variability in the photospheric line shapes, and the systemic velocity is often measured using the narrow disk line components. For the systemic velocity, we adopt a value of 20.18 km/s as the average of Fe\,I disk lines measured recently by \citet{Kiefer2019}.\\

\begin{figure*}
    \centering
    \includegraphics[width=\textwidth,trim={6cm 0.5cm 6cm 2cm},clip]{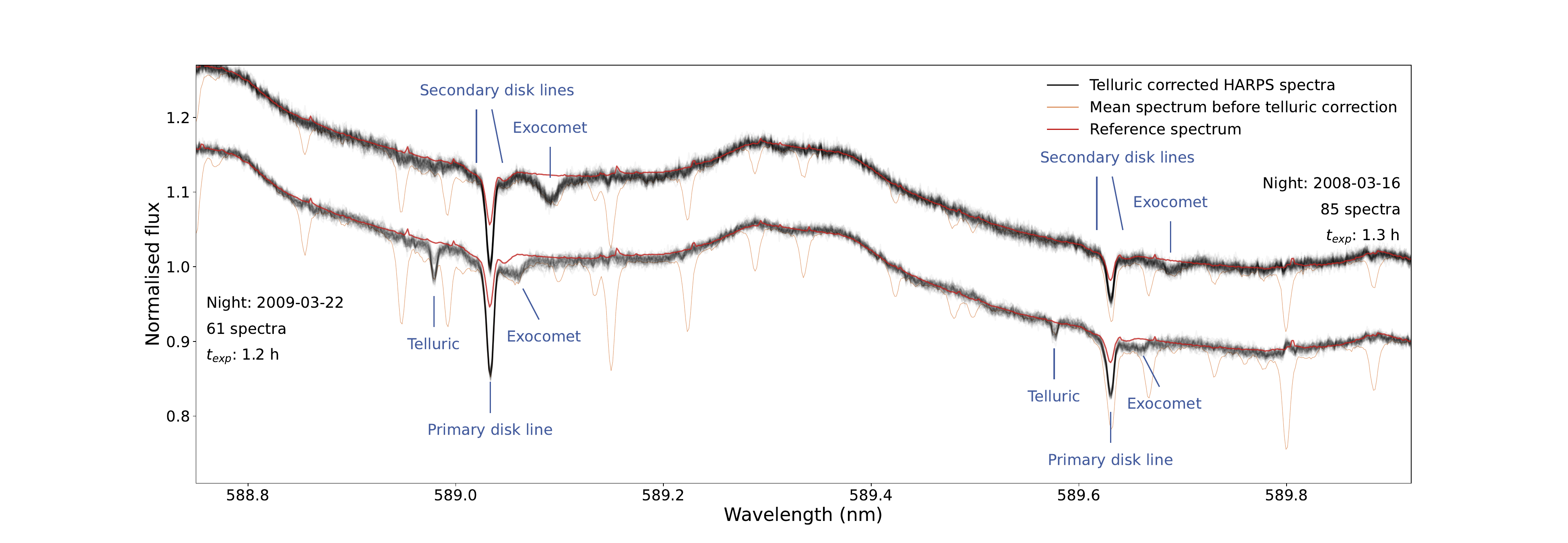}
    \caption{Telluric-corrected HARPS observations at the sodium doublet, obtained on two nights nearly one year apart, vertically offset for clarity. In both nights, significant disk line absorption is visible, as well as red-shifted lines attributed to a form of exocomet activity. Labels in black indicate the nights of observation, the number of spectra taken and the total exposure time. The night-average spectrum prior to telluric correction is over-plotted in orange, demonstrating the relative positions of the telluric water lines.}
    \label{fig:Na}
\end{figure*}

\section{Results}
\subsection{Rare exocometary Na\,I absorption}
After telluric correction we find a general absence of strong, variable Na\,I absorption similar to the strongly absorbing stochastic activity in the Ca\,II lines. Out of 198 nights of observation, we identify only two nights in which clear additional ($> \sim 2\%$) sodium absorption is present in both lines of the Na\,I doublet, in the nights of March 16 2008 and March 22 2009, strongly red-shifted from the central disk line (see Fig. \ref{fig:Na}). In addition, we discern red-shifted Na\,I absorption during an epoch spanning approximately 13 nights in February 2004, where it is present persistently with a depth of approximately $1\%$ (see below).\\ 

\noindent In addition, we find strong, narrow Na\,I absorption at blue-shifts between 12 and 28 km/s away from the Na\,I disk lines in approximately one third of the observed nights. These velocities coincide with the barycentric velocity at the time of observation (see Fig. \ref{fig:Na_2} for data obtained in 2017-2018), and we identify this as uncorrected telluric sodium absorption at the $1\%$ level.\\

\begin{figure*}
    \centering
    \includegraphics[width=\textwidth,trim={0cm 4cm 0.2cm 0.2cm},clip]{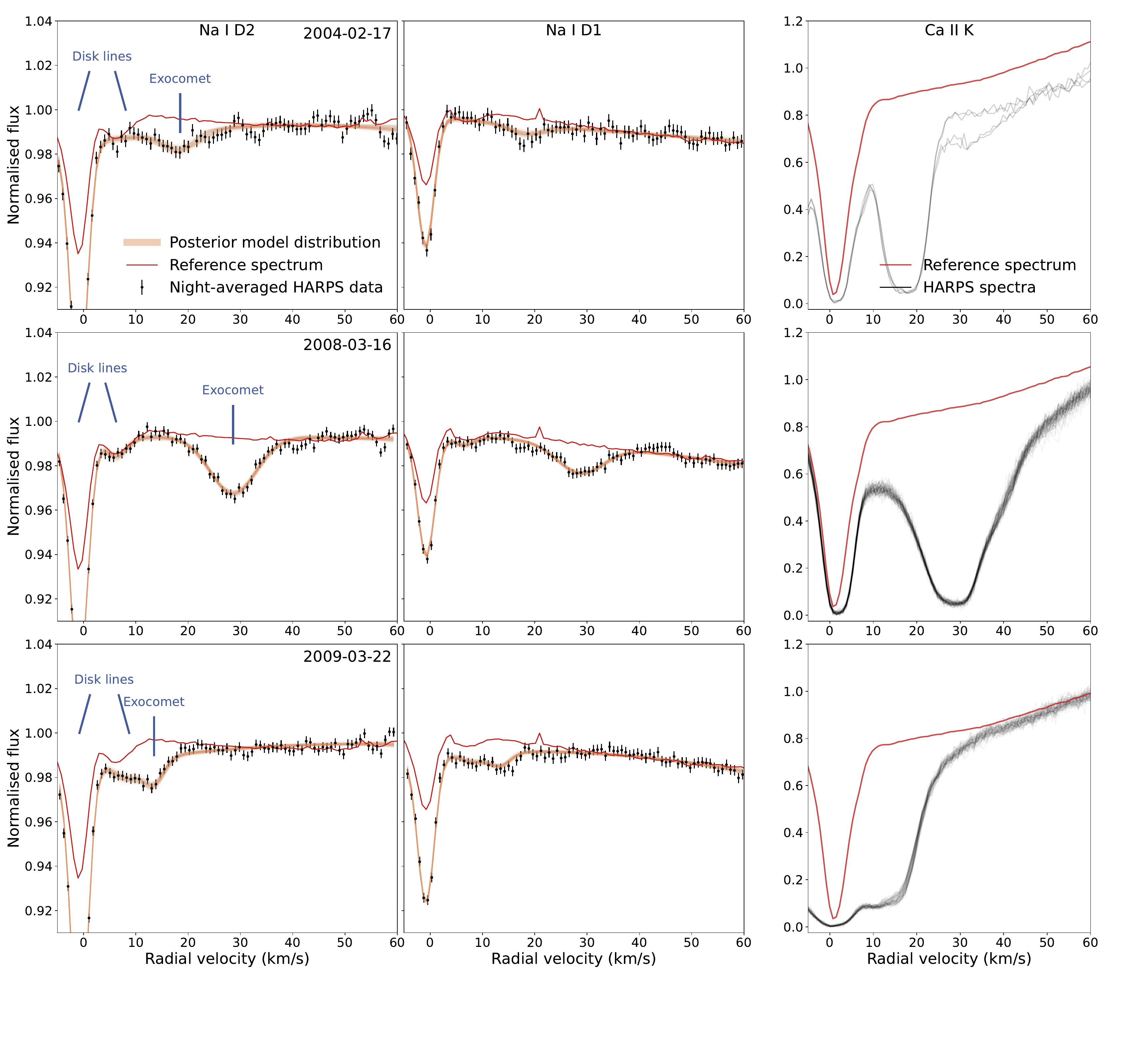}
    \caption{Best-fit models to the Na\,I D lines (left and middle columns) on three nights of data (top, middle and bottom rows), compared to the Ca\,II K line (right column). The night of March 17 2008 (middle row) shows the strongest absorption of exocometary sodium of all 198 nights of data. The night of February 17 2004 is part of a sequence of 13 nights in which weak exocometary sodium absorption persists, and this is the night with the best detectability of this sequence but the weakest exocometary signal detected in the entire dataset. The posterior parameter distributions of this fit are shown in Fig. \ref{fig:Na_cornerplot} as a representative example. In all three cases, the exocometary sodium absorption occurs simultaneously with strong Ca\,II absorption at the same relative radial velocity -- indicating that the sodium is sublimated together with calcium by the same exocomets.}
    \label{fig:Na_linefits}
\end{figure*}

\noindent We use our fitting method (see Methods) to fit the two exocomet features of March 2008 and March 2009, as well as one epoch from the sequence of February 2004 in which the weak exocomet absorption is most clearly visible. The resulting fits are shown in Fig. \ref{fig:Na_linefits}, and compared with the Ca\,II K-line at the same epochs. In each fit, we include the Na\,I disk line as two blended components, a strong central line with a secondary red or blue-shifted line (see Fig. \ref{fig:Na}). Because both the disk lines and the variable exocomet features are observed to have line depth ratios approximately equal to two, we infer the absorption to be unsaturated. At the same time, the lines are  relatively weak, with line-depths on the order of 1\% of the stellar flux. The optical depth of the absorbing cloud must thus be small $(\tau_0 \ll 1)$. In this regime, $f$ and $\tau_0$ are strongly degenerate and this prevents us from determining confidently whether the exocometary Na\,I clouds fill the entire stellar disk. For this reason we fix the fill fraction $f$ to 100\%, making $\tau_0$ essentially a lower limit on the true optical depth, and a proxy for the line width $D$ (see above), which is tightly constrained in all three cases. In the 2004 sequence, the D1 line might actually be deeper than half of the depth of the D2 line (see Fig. \ref{fig:Na_linefits}), which would indicate mild saturation. Leaving the fill fraction $f$ free leads the algorithm to prefer small fill fractions, but these are only inconsistent with $100\%$ at a level between 3 and 4 $\sigma$, so we consider this constraint to be only tentative and continue to work with effective line depths in the remainder. We also systematically fit the blue-shifted telluric features, where it is clearly visible by eye in 71 of the 198 nights of observation, and constrain the distribution of telluric sodium line depths (see Fig. \ref{fig:Na_distributions}), demonstrating that the Earth's atmosphere routinely contributes absorption at the 1\% to 5\% level.\\

\noindent All fits are carried out with the same priors (see Table \ref{tab:priors_Na}), apart from the prior on the center wavelength of the target component, which we shift according to its approximate blue or red-shift. An example of the posterior distributions of the model parameters of the exocometary Na\,I lines in the night of February 17, 2004, which is the weakest signal that we decided to fit. The line depth is constrained to $0.89\pm0.13\%$, and thus the line is detected with high confidence. A comparison between the wavelengths of the Na\,I and the Ca\,II absorption lines reveals that exceptionally deep exocometary Ca\,II are present at the rare times during which significant Na\,I is detectable, at the same radial velocity (see Fig. \ref{fig:Na_linefits}). Although more detections of Na\,I would be needed to prove a more statistically robust correlation, we infer that the Na\,I we detect is sourced from the same exocometary bodies that are producing large clouds of Ca\,II. It is not unexpected that rocky exocometary material would source both calcium and sodium, but because of the short photo-ionization timescale of sodium, its column density may be expected to be insignificant. Indeed, most calcium absorption events are not associated with significant Na\,I absorption. The present observations suggest that Na\,I can still be detectable during episodes of strong Ca\,II production. \\

\noindent The observation of lines of multiple species can in theory be used to determine elemental ratios of the evaporating exocometary material, and this has recently been convincingly demonstrated by \citet{Vrignaud2025}. However, we expect that in the case of neutral sodium, the simple models used until now are unlikely to be able to provide accurate abundance measurements. In this work and in previous literature \citep[e.g.][]{Vrignaud2025}, exocomet clouds have been modeled as obscuring the star homogeneously, while in reality they may exhibit radially varying density profiles, depending on their rates and balances of ionization and their dynamics. As Na\,I is very easy to ionize, we expect it to be present only close to the nucleus (potentially consistent with low values for $f$ tentatively found earlier), while Ca\,II may be present much further away. In cases of very strong outgassing, there may be regions that are shielded from ionizing photons, and this may explain why Na\,I occurs simultaneously with very deep Ca\,II events). We conclude that determining the Na\,I / Ca\,II ratio may be possible, but that this would require more detailed physical modeling of the particle dynamics, resulting morphology and radiative properties of the exocometary cloud.\\

\begin{figure}
    \centering
    \includegraphics[width=0.5 \textwidth,trim={0.2cm 0.2cm 0.5cm 1.0cm},clip]{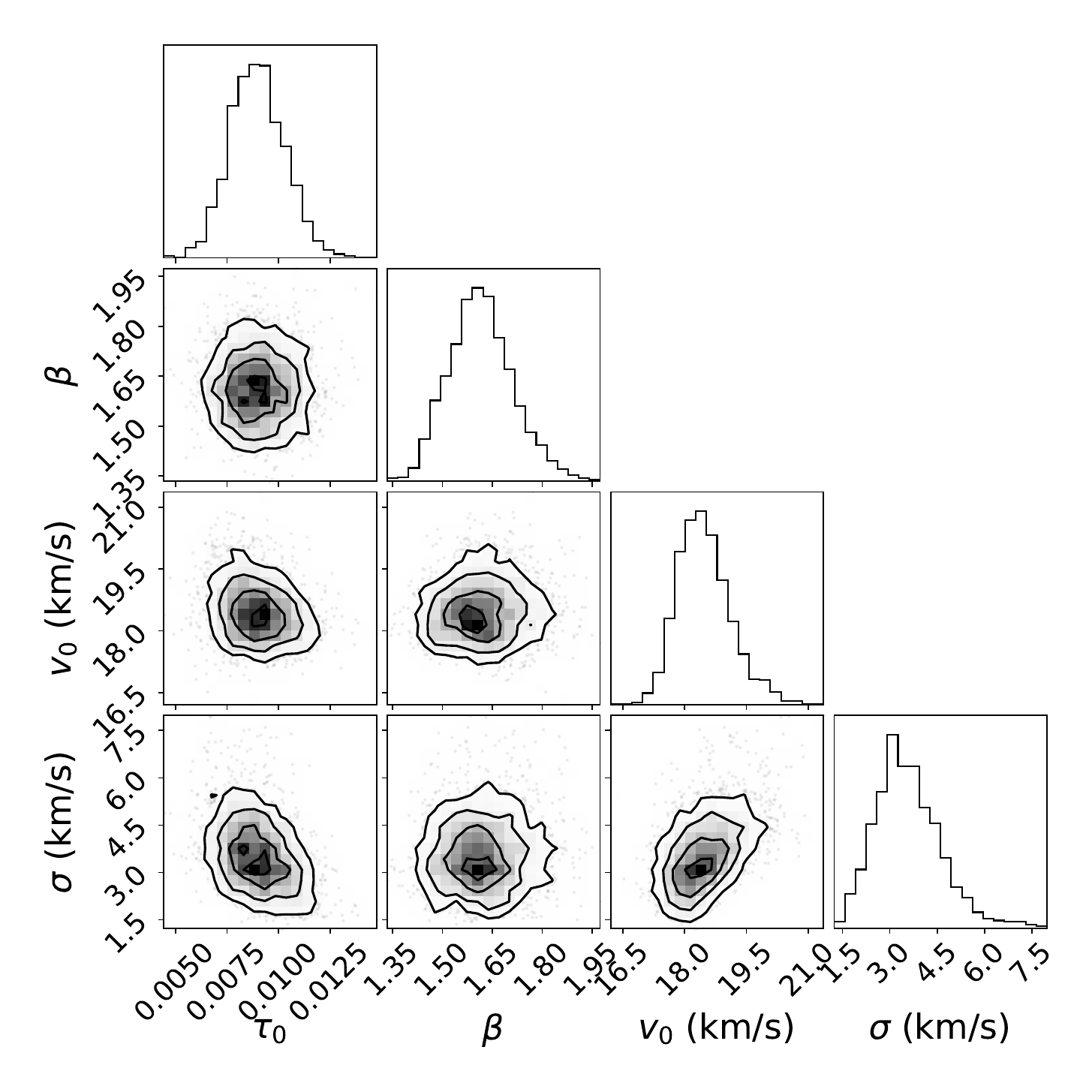}
    \caption{Posterior distributions of the parameters that describe the exocometary sodium lines on the night of February 17, 2004 shown in Fig. \ref{fig:Na_linefits}. Because the observed Na\,I lines are highly optically thin, $\tau_0$ and $f$ are were not independently constrained. Instead, we fixed $f$ to 1 and converted $\tau_0$ into line depth $D$, which is constrained to $0.89\pm0.13\%$. Note that $\beta$ may be elevated due to unmodeled variance in the spectrum caused by residuals from the telluric correction process.}
    \label{fig:Na_cornerplot}
\end{figure}

\subsection{Two anomalous exocomet events}
\begin{figure}
    \centering
    \includegraphics[width=0.5\textwidth,trim={2cm 1cm 6cm 2cm},clip]{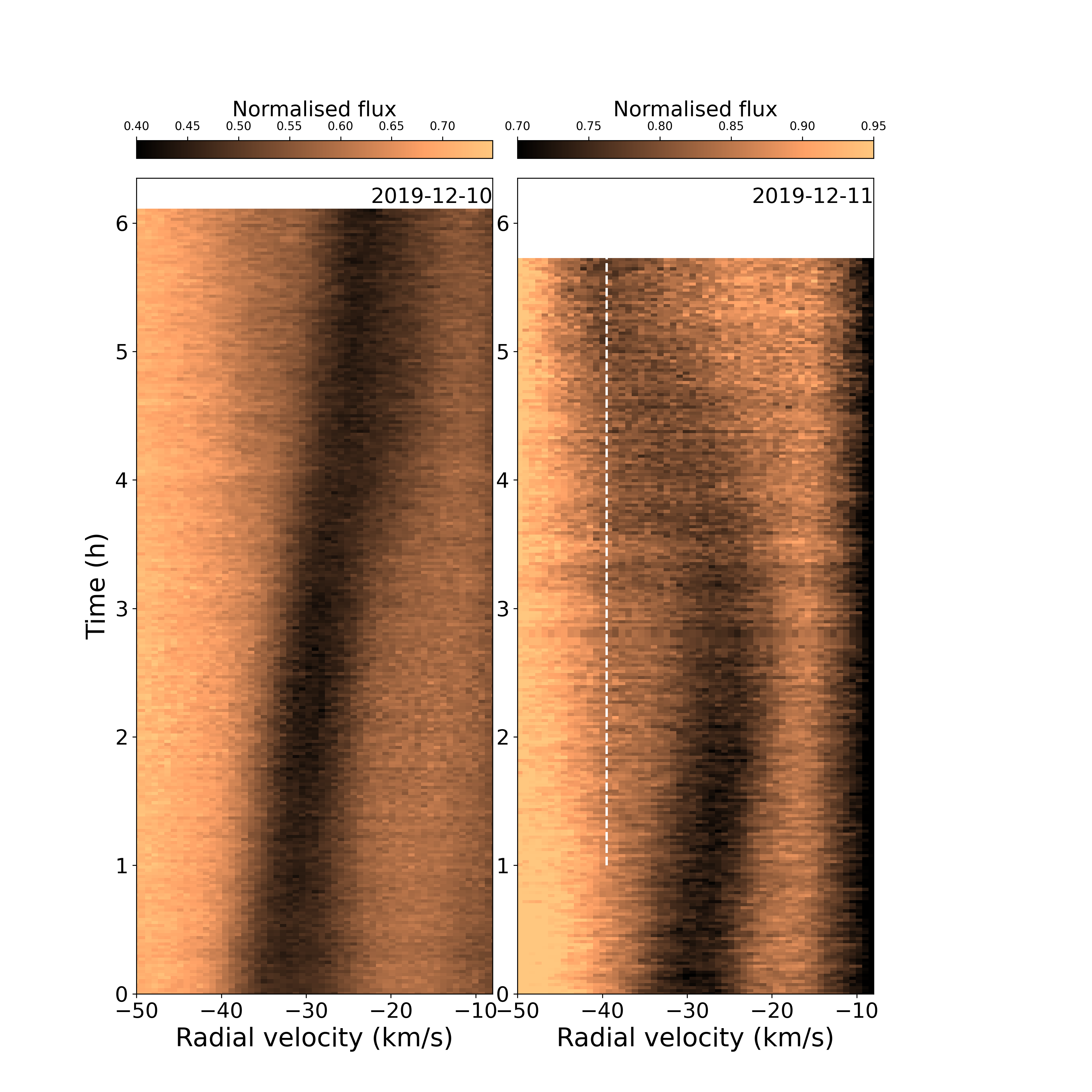}
    \caption{Spectroscopic time-series of two nights of observations blue-wards of the Ca\,II K-line\, calibrated using the reference spectrum.  The observing sequences both lasted for about 6 hours, and took place 24 hours apart. The disk line is near 0 km/s, and a strong blue-shifted exocomet feature absorbs near -30 km/s on both nights that is decelerating towards smaller blue-shifts. In the second night, a third component appears statically near -40 km/s that blends with the decelerating feature. This component is indicated with the dashed line. The line parameters of these exocomets are shown in Fig. \ref{fig:linefits}}
    \label{fig:Ca2D}
\end{figure}

On two sequential nights of observation in 2019 (program ID 0104.C-0418, P.I. Lagrange, on 11 and 12 December 2019, hereafter nights 1 and 2), strongly blue-shifted exocomet signatures were observed in the Ca\,II H\&K lines, that experience significant acceleration during their transit (see Fig. \ref{fig:Ca2D}), and occur at similar radial velocities. Assuming that their orbital periods are much greater than 24 hours (see below) we conclude that these are two objects on the same 'comet train' on a similar orbit, transiting nearly 24 hours apart.\\

\noindent Observations of accelerating exocomets are relatively rare \citep{Ferlet1987,Kennedy2018}, as these require observations spanning multiple hours and a comet that is very close to the star for its orbital curvature to cause significant acceleration. Accelerating blue-shifted features are especially rare because exocomets are mostly observed while moving towards the star, away from the observer \citep{Kiefer2014}. Blue-shifted exocomets are significant because they are retreating from the star, implying that they have survived at least one close-in periastron passage, and because such a trajectory is only a rare outcome of the mean-motion resonance formation model \citep{Beust2024}, meaning that these comets may instead have a different dynamical origin (see Jaworska et al. in prep. for more discussion about the dynamics of blue-shifted exocomets) from the population that is observed predominantly at red-shifted velocities. Previous efforts to determine the statistics of orbital elements have been hampered by the fact that most exocomet absorption lines are stable on $\sim 1$ hour time-scales, and a singular radial velocity measurement leaves strong degeneracies in possible orbital parameters. Observable acceleration of a transiting exocomet allows for some of this degeneracy to be lifted \citep{Kennedy2018}. \\

\begin{figure}
    \centering
    \includegraphics[width=0.5\textwidth,trim={0.5cm 1.5cm 1cm 3cm},clip]{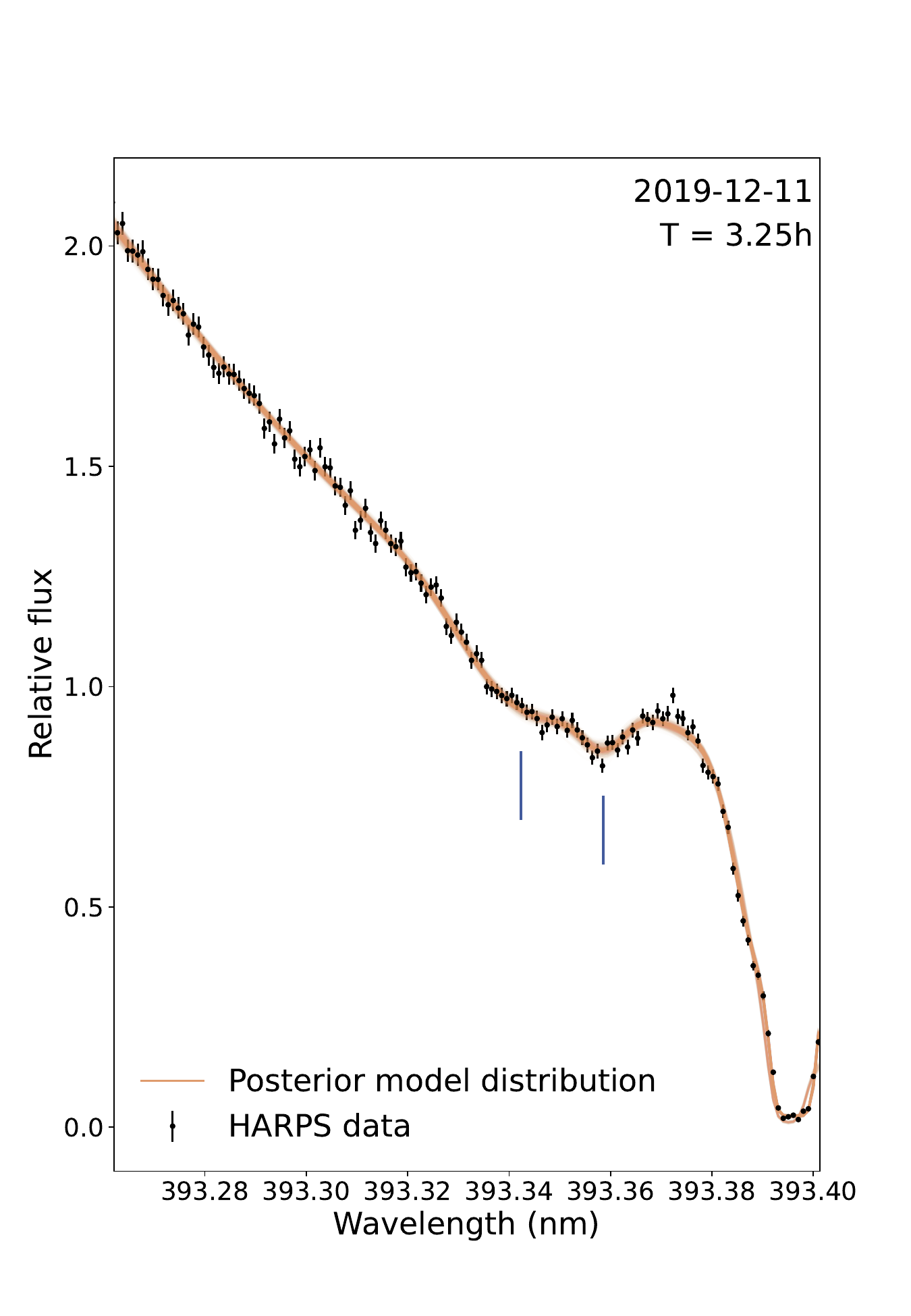}
    \caption{Example of a fit of the blended components during night 2, indicated by the vertical blue markers. The fit confidently distinguishes the two components.}
    \label{fig:fitting_example2}
\end{figure}

\begin{figure}
    \centering
    \includegraphics[width=0.5\textwidth,trim={1cm 2cm 3cm 3cm},clip]{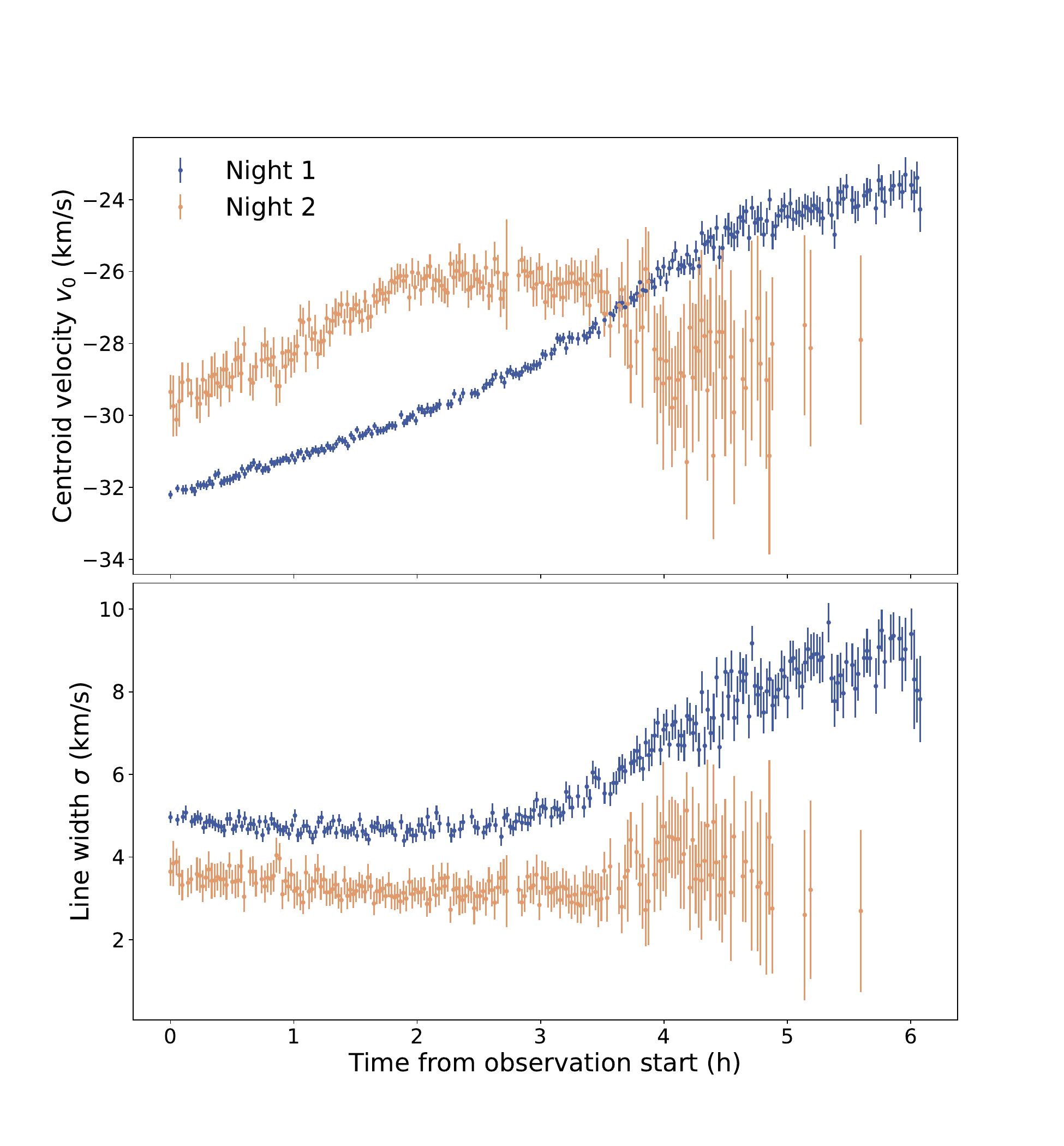}
    \caption{Centroid velocities (top panel) and Gaussian line widths (bottom panel) of the exocomet features shown in Fig. \ref{fig:Ca2D}. Both bodies follow a near-constant acceleration at the start of the observations, as expected from a Keplerian orbit (see Fig. \ref{fig:paramspace}). However, they then deviate from Keplerian acceleration towards the second half of the observing sequence. Simultaneously, the observed line width increases dramatically in the first night, and may increase in the second night as well. In some cases, in particular towards the end of night 2 when the two components blend, the fit was unstable resulting in prior-bound parameter estimates. These points are omitted.}
    \label{fig:linefits}
\end{figure}

\noindent We proceeded to use our line-fitting algorithm (see Methods) to determine the radial velocities and line widths of the accelerating features in both nights, with priors shown in Table \ref{tab:priors_Ca}. Singular examples of these fits were already shown in Figures \ref{fig:cornerplot} and \ref{fig:fitting_example}. On the second night of observation, a clear signature of a static or weakly accelerating blue-shifted exocomet appears in the vicinity of -39 km/s, which partially blends with the stronger accelerating feature, overlapping with it towards the end of the observations (see Fig. \ref{fig:Ca2D}). To fit the accelerating exocomet in the second night reliably, we therefore add a second exocomet component in the fit with priors included in \ref{tab:priors_Ca}, with a typical result shown in Fig. \ref{fig:fitting_example2}. Best-fit parameters for both nights are plotted in Fig. \ref{fig:linefits}, and we note that the uncertainty on fitting parameters increases at the end of the second night because of increased blending between the two components. Peak optical depths $\tau$ were typically found the be close to unity, with fill factors near 0.5 and consistent with full coverage of the stellar disk in some of the spectra, though typically with $\sim 10\%$ uncertainty. These properties may be anomalous compared to the exocomet families distinguished by \citet{Kiefer2014}, who showed that small line-widths and large fill factors are generally associated with exocomets with small positive radial velocities. The features detected here have relatively large fill factors, consistent with D-family comets, but also strong negative blue-shift consistent with S-family comets.\\  

\begin{figure}
    \centering
    \includegraphics[width=0.5\textwidth,trim={7cm 3cm 7cm 2cm},clip]{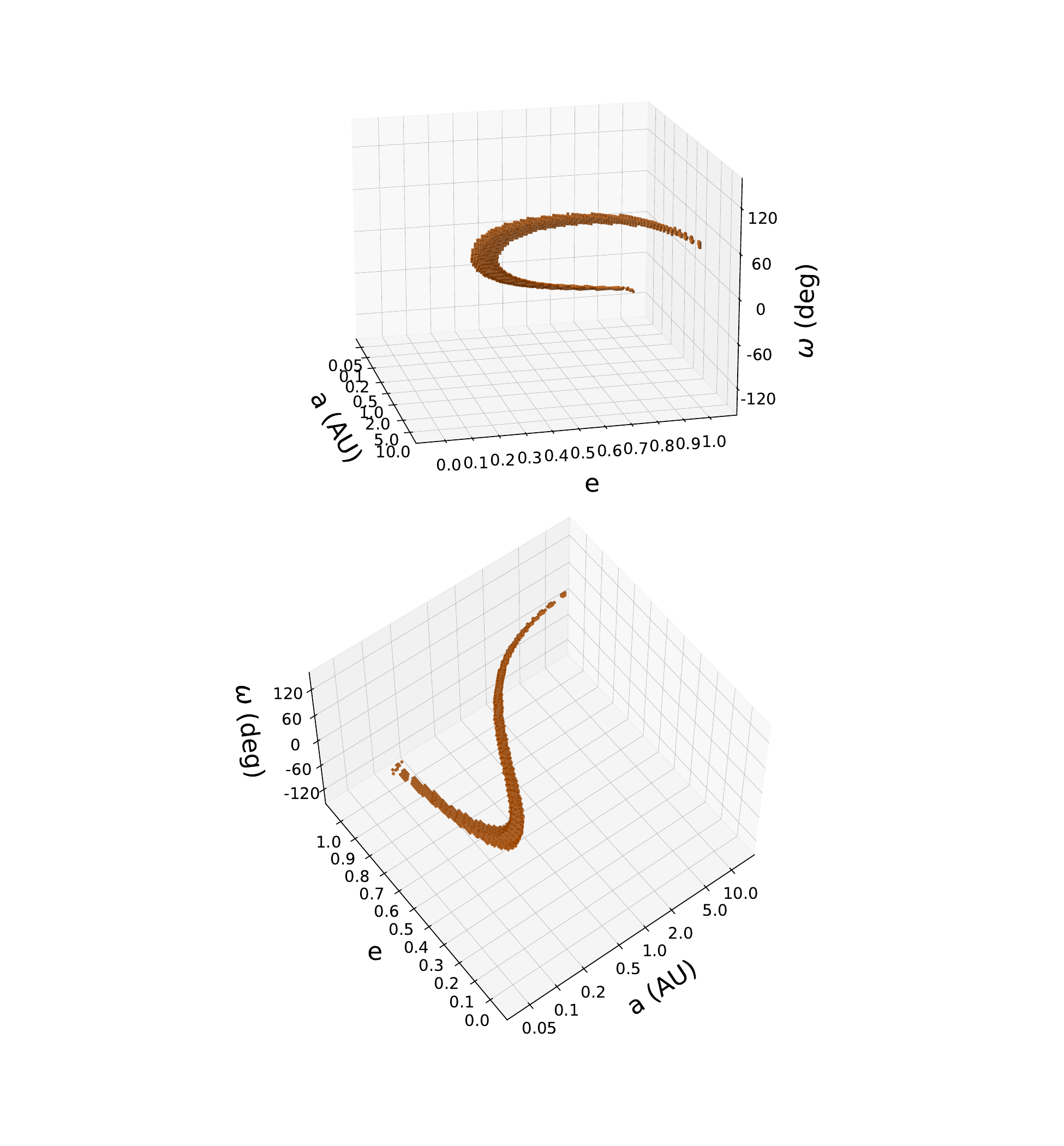}
    \caption{Family of orbital solutions that have a mean radial velocity during the transit between -35 and -27.5 km/s and a mean radial acceleration between 0.9 and 1.3 km/s/h, consistent with the linear part of the exocomet radial velocity observed during the first three hours of observation in the first 2019 night, and assuming an inclination $i$ equal to 0.0 and a fixed longitude of ascending node $\Omega$ (degenerate with argument of periastron $\omega$ when $i=0$). A linearly accelerating absorption line still permits a wide combination of orbital eccentricities, semi-major axes and orbital orientations.}
    \label{fig:paramspace}
\end{figure}

 \noindent On both nights of observation, the absorption lines are accelerating linearly near -32 km/s for the first three hours and two hours respectively. This is expected from a Keplerian orbit, because the radial velocity is to first order a narrow linear segment of a pseudo-sinusoid, covering a few hours of an orbit with a much longer period \citep{Kennedy2018}. On the first night, we determine that this acceleration is nearly 1.1 km/s/h. To determine which class(es) of the orbit are consistent with these velocities (initially chosen in preparation for defining Bayesian fitting priors), we compute the mean in-transit velocity of a three-dimensional space varying semi-major axis $a$, eccentricity $e$ and argument of periastron $\omega$, leaving the inclination $i$ fixed to 0 (because this is a transiting geometry) and fixing the longitude of ascending node $\Omega$ to 180 degrees, as it is fully degenerate with $\omega$ in this geometry. We then determine which orbits have a mean radial velocity during transit between -27.5 and -35 km/s, while experiencing an acceleration between 0.9 and 1.2 km/s/h. The parameter combinations satisfying these criteria are shown in Fig. \ref{fig:paramspace}, revealing that a wide range of orbits with $e>0.3$ could be consistent with the observed motion. Not shown in Fig. \ref{fig:paramspace} is a matching class of orbits of arbitrary eccentricity, where the periastron is along the line-of-sight ($\omega = 90^{\circ}$), if only the first half of the transit (blue-shifted) is covered by the observations. Also not shown are hyperbolic orbits. This serves to illustrate that robustly determining the orbital parameters of these objects is not possible without additional data or assumptions. A common assumption is to adopt a parabolic orbit, which removes two degrees of freedom from the problem. However, recent dynamical simulations by \citet{Beust2024} place the origin of exocomet progenitors within 1 AU from the star. Given that Ca\,II sublimation sets on within approximately 0.4 AU \citep{Beust1998}, assuming very high eccentricity trajectories may not be justified. In addition, we remark that such fitting procedures assume that the observed cloud of calcium follows the Keplerian velocity of the sourcing body, while in reality the ion tail should be expected to be blown in the anti-stellar direction, introducing an additional and unknown (range of) blue-shift(s).\\

\noindent Instead of attempting to further constrain the orbital parameters of the transiting exocomet features in the two nights of 2019 data, we turn our attention to the clear departures from Keplerian velocities occurring later during the observations. In the fourth hour of observation on the first night, the acceleration of the comet increases, before decreasing one hour later. This behavior cannot be reconciled with Keplerian motion. More dramatic is the acceleration of the body in the second night -- which is presumably on the same orbit as the first body if both are on the same 'comet train' nearly 24 hours apart. After two hours into the observation, the acceleration of this object appears to reverse, before fully blending with the more static second component at the end of the sequence.

\noindent An apparent blue-ward acceleration of an exocomet may be expected if the observations cover the transit egress. In this scenario, the leading part of the cometary tail is removed from view, while the rest remains in-transit. Since material further along the tail is presumably more blue-shifted due to radiation pressure than closer to the nucleus, this could cause an apparent blue-shift of the disk-integrated absorption line. However, this scenario would predict that in the early part of the observations, the line would be wide, before narrowing as the least blue-shifted components egress. In these observations (particularly in night 1), the opposite happens. Secondly, the reversal of the acceleration in the second night is sudden, taking place in approximately an hour, while extended clouds have ingress/egress times of several hours. Finally, we note that the behavior of both objects is different. Assuming that these objects are indeed moving on the same orbit, their behavior during egress could be expected to be similar.\\

The presence of the blended component in night 2 complicates the fit, and this is evident from the increased uncertainties on the best-fit parameters. In theory, the presence of weaker, blended exocomet features on similar or different orbits could act to skew the centroid velocities, but we do not see clear evidence of additional absorbing features and note that blue-shifted exocomets are altogether relatively rare \citep{Kiefer2014}.

\noindent Instead, we propose an alternative scenario to interpret the sudden departure of Keplerian motion seen in Night 2 as providing evidence for the final fragmentation of the exocomets' nucleus and subsequent blow-out of the cometary tail, briefly after periastron passage. Destruction of stargrazing comets may be sudden if the core of the nucleus contains unexposed, previously unevaporated volatiles (ice) that evaporate in response to the intense heating during periastron passage. This evaporation may be delayed until after periastron passage, as the heat pulse first needs to propagate through the rocky exterior before evaporating the icy interior. This scenario is comparable to Kreutz-family Sungrazers in the solar system, for example the post-periastron destruction of the nucleus of comet Lovejoy in 2011 \citep{Sekanina2012}. As this scenario may favor an icy interior composition, this further strengthens the possibility that the origin of these two exocomets is not among the more commonly observed red-shifted exocomets presumed to be generated from the devolitalised inner planetesimal belt via mean-motion resonance with \betapic c, but instead may have originated from the outer system exterior to the snow-line relatively recently. \citet{RodetLai2024} as well as our companion paper (Jarworska et al. in prep) suggest that secular evolution of icy exocomet progenitors from the outer system is a viable pathway to producing observable exocomets. Our dynamical simulations (Jaworska et al. in prep) indicate that exocomet progenitors likely spend less than $10^2$ to $10^3$ years within the water sublimation limit before reaching within 0.4 AU of the star where they are spectroscopically observed as exocomets. Given a sufficiently large nucleus, an icy core may survive throughout the comets' passage through the inner system before becoming an exocomet, and this has been explored previously by \citet{Karmann2001}.\\

\subsection{Coherence of exocomet lines and disk line variability}
As previously shown for a subset of this data \citep{Kiefer2014}, exocometary Ca\,II absorption is extremely variable and occurs over a wide range of (mostly red-shifted) radial velocities up to 100 km/s. Some exocomet absorption lines vary on timescales of hours, while some absorption signatures are coherently present for days. Variability on hour timescales is naturally expected from transiting objects on eccentric orbits at close distances to the star, while longer-term coherence supports the notion of 'comet families': trains of objects on similar orbits but separated from one another in time. This is naturally expected from a formation scenario in the mean-motion resonance model \citep{Beust2024}. However, we hypothesize that planetesimals from the inner system that turn into exocomets via mean-motion resonance with \betapic c may make close passes with the planet and be susceptible to tidal disruption. Tidal disruption may also explain the prevalence of long timescale coherence in the observed Ca\,II exocomet signatures  of planetesimals originating from the outer system, as these typically undergo chaotic dynamical evolution after being scattered into the inner system.\\

\noindent However, the data also provide examples of much longer time-stability: From early 2017 to mid-2018, there is a $\sim 5$ km/s blue-shifted feature that appears to last continuously (see Fig. \ref{fig:Ca_coherence}). Dynamical simulations should be carried out to investigate whether such long coherence timescales are readily explained by the exocomet model, or whether this feature is due to some other component of the circumstellar material.\\

\noindent Finally, from investigating the large volume of observations at the Ca\,II lines by eye, and by analyzing the best-fit line parameters of the Na\,I D-lines, we conclude that the disk lines show significant variability that may hinder the usage of percentile-based reference spectra \citep[as per][]{Kiefer2014}. Figure \ref{fig:Na_disk_distributions} shows that the relative line-depth of the Na\,I disk lines varies by more than a factor of two from $\sim 5\%$ up to $13\%$. The line center varies by a few hundred meters per second, with a mean measurement uncertainty of 50 m/s. Although the Ca\,II doublet appears to be more stable, it also shows variability. Fig. \ref{fig:CaII_variability} shows an example of a night of observation in 2014 where the observed disk line is significantly narrower and shifted compared to the reference spectrum (see Methods). \citet{Kiefer2019} also observed variability in the Fe\,I disk line, indicating that the tenuous circumstellar gas is generally observably variable.

\section{Conclusions}
In this paper we investigated the growing archive of HARPS observations of the \betapic system in search for new exocomet phenomenology. This paper is the first in a series that seeks to investigate the nature of the exocomets of \betapic, leveraging the large number of high-quality spectra obtained for nearly two decades. We started by focusing on region of the sodium D-line doublet, that overlaps with significant telluric water absorption. We systematically corrected these spectra using \texttt{Molecfit} and that exocometary sodium absorption is relatively rare, with only three distinct epochs showing detectable signals. These events are contemporaneous with exceptionally strong Ca\,II absorption occurring at the same relative radial velocity, and we conclude that the observed sodium and calcium components are sourced by the same bodies. We also investigated two anomalous exocomet events at the Ca\,II lines observed in 2018, which show a strong and sudden onset of non-Keplerian acceleration that we hypothesize to be evidence for the destructive fragmentation of the exocomet nucleus and the dispersal of the Ca\,II tail. Finally, after creating a qualitative overview of the phenomenology in the \betapic spectra but without doing a thorough statistical analysis, we conclude that there is significant disk line variability in Ca\,II and Na\,I, as well as long timescale coherence in some presumed exocomet absorption signals that as yet remain unexplained.

\begin{acknowledgements}
This work was supported by grants from eSSENCE (grant number eSSENCE@LU 9:3), the Swedish National Research Council (project number 2023-05307), The Crafoord foundation and the Royal Physiographic Society of Lund, through The Fund of the Walter Gyllenberg Foundation. We are grateful to Brett Morris for useful discussions regarding the inference method used extensively in this work, and the attendees of the 2024 ISSI workshop on Exocomets in Bern, Switzerland, for important discussions regarding the telluric Na\,I lines. We are also grateful by the insightful comments and advice provided by an anonymous referee, who is to be credited for remarking the existence of the blended component in the second night of the 2019 observations.
\end{acknowledgements}

%
  \bibliographystyle{aa} 
  \bibliography{references} 
%

\begin{appendix} 
\section{Additional tables and figures}

\begin{figure}[!htb]
    \centering
    \includegraphics[width=0.5\textwidth,trim={1.2cm 0.7cm 2.5cm 2.5cm},clip]{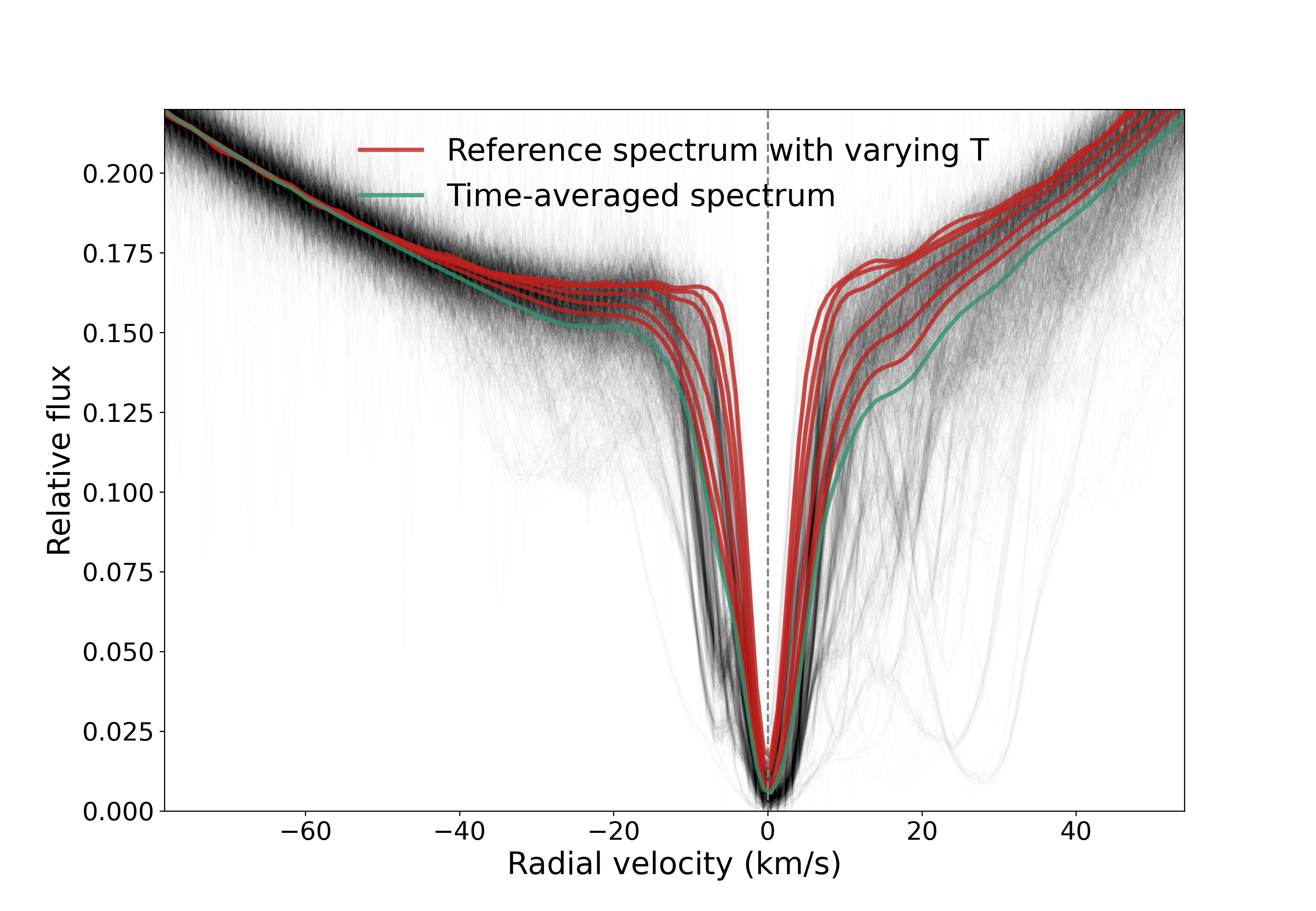}
    \caption{Comparison of the reference spectrum (red) computed for varying values of the threshold $T$ as the 98$^{\mathrm{th}}$, 90$^{\mathrm{th}}$, 80$^{\mathrm{th}}$, 50$^{\mathrm{th}}$, 25$^{\mathrm{th}}$ or 10$^{\mathrm{th}}$ percentile (red lines from top to bottom). The reference spectrum obtained by setting the threshold to the 80$^{\mathrm{th}}$ lies within 2.5\% of the reference spectrum obtained using the 98$^{\mathrm{th}}$ percentile at all velocities greater than $\pm$ 15 km/s from the center of the disk line. The time-average spectrum is shown in green, corresponding to the 0$^{\mathrm{th}}$ percentile reference spectrum.}
    \label{fig:percentiles}
\end{figure}

\begin{figure}
    \centering
    \includegraphics[width=0.5 \textwidth,trim={2cm 1cm 3cm 2.5cm},clip]{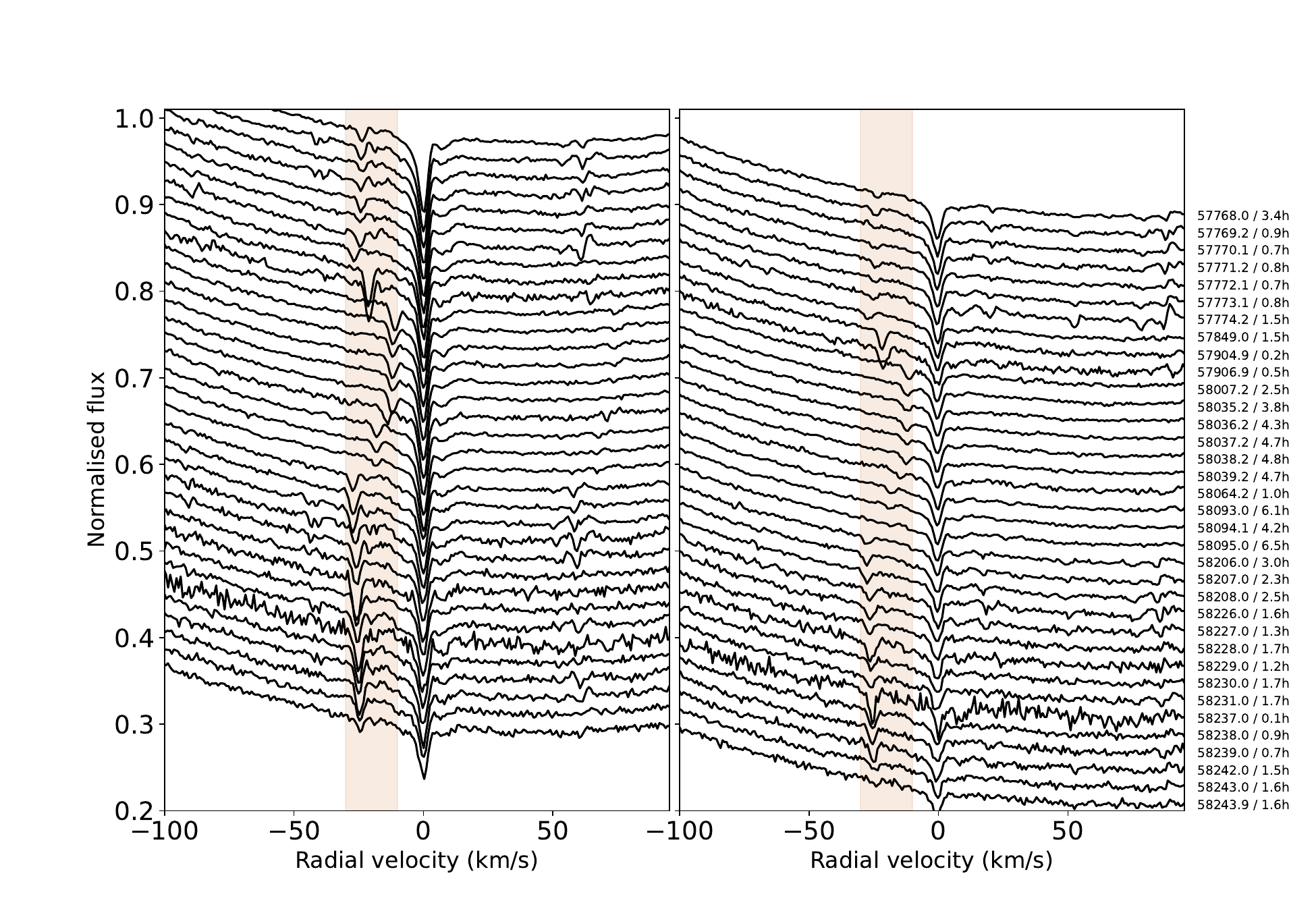}
    \caption{Nightly averages of telluric-corrected spectra at the sodium doublet, vertically offset for clarity. These spectra are obtained between January 2017 to June 2018 and show how a telluric Na\,I feature is continuously present with a radial velocity varying within the range of radial velocities of the Earth around the barycenter (shaded orange).}
    \label{fig:Na_2}.
\end{figure}

\begin{figure}
    \centering
    \includegraphics[width=0.5 \textwidth,trim={1.7cm 5cm 2.2cm 40cm},clip]{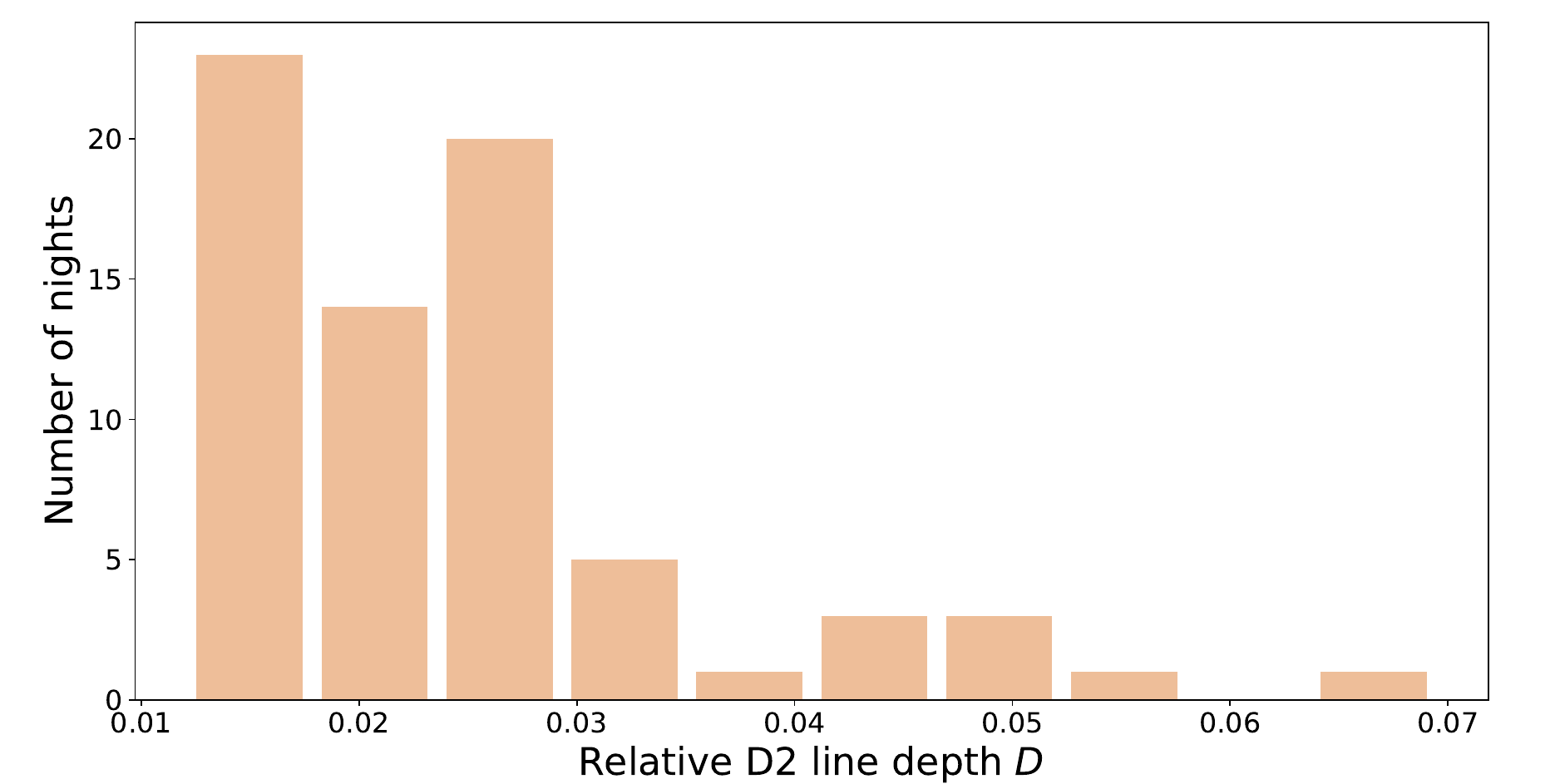}
    \caption{Best-fit depths $(D)$ of the telluric D2 line, for all 71 nights in which telluric Na I was discerned.}
    \label{fig:Na_distributions}
\end{figure}

\begin{table}
  \caption[]{Priors used when fitting exocometary sodium in four nights (see Fig. \ref{fig:Na_linefits}), describing three line components occurring in two wavelength slices.}\label{tab:priors_Na}
\begin{center}
\begin{tabular}{lll}
\hline
\hline
Parameter & Prior & Comment \\
\hline
$\beta$ & $U\left( 0.8,10.0 \right)$ &  \\
$\tau_1$ & $U\left( 0.0,3.0 \right)$ & Main disk line component. \\
$\tau_2$ & $U\left( 0.0,3.0 \right)$ & Blue-shifted component. \\
$\tau_3$ & $U\left( 0.0,2.0 \right)$ & Secondary disk line. \\
$\lambda_1$ & $U\left( 589.03,589.04 \right)$ &  \\
$\lambda_2$ & variable &  \\
$\lambda_3$ & $U\left( 589.04,589.06 \right)$ &  \\
$\sigma_1$ & $U\left( 0.8,4.0 \right)$ &  \\
$\sigma_2$ & $U\left( 0.8,4.0 \right)$ &  \\
$\sigma_3$ & $U\left( 2.0,5.0 \right)$ &  \\
$f_1$ & 1 & Fill fractions fixed. \\
$f_2$ & 1 &  \\
$f_3$ & 1 &  \\
$d \lambda$ & $U\left( 0.59,0.605 \right)$ &  \\
$c_0$ & $U\left( 0.95,1.05 \right)$ & Continuum D2 line. \\
$c_1$ & $U\left( -0.01,0.01 \right)$ &  \\
$c_2$ & $U\left( -0.005,0.005 \right)$ &  \\
$d_0$ & $U\left( 0.9,1.1 \right)$ & Continuum D1 line. \\
$d_1$ & $U\left( -0.05,0.05 \right)$ &  \\
$d_2$ & $U\left( -0.005,0.005 \right)$ &  \\
\end{tabular}
\end{center}
\end{table}

\begin{table}
  \caption[]{Priors used when fitting the Ca II line doublet in the two 2019 nights, describing three line components occurring in four wavelength slices. The priors on polynomial coefficients $c_i$ are the same for all four slices and are not repeated in the table.}\label{tab:priors_Ca}
\begin{center}
\begin{tabular}{lll}
\hline
\hline
Parameter & Prior & Comment \\
\hline
$\beta$ & $U\left( 0.5,3.0 \right)$ &  \\
$\tau_1$ & $U\left( 4.0,15.0 \right)$ &  Main disk line component.\\
$\tau_2$ & $U\left( 0.0,3.0 \right)$ &  Exocomet.\\
$\tau_3$ & $U\left( 0.0,4.0 \right)$ &  Secondary disk line. \\
$\tau_4$ & $U\left( 0.0, 2.0 \right)$ & Blended line (only night 2). \\
$\lambda_1$ (nm) & $U\left( 393.390,393.410 \right)$ & H-line center. \\
$\lambda_2$ (nm) & $U\left( 393.347,393.370 \right)$ &  \\
$\lambda_3$ (nm) & $U\left( 393.380,393.395 \right)$ &  \\
$\lambda_4$ (nm) & $U\left( 393.341,393.346 \right)$ &  \\
$\sigma_1$ (km/s) & $U\left( 1.5,7.0 \right)$ &  \\
$\sigma_2$ (km/s) & $U\left( 2.0,12.0 \right)$ &  \\
$\sigma_3$ (km/s) & $U\left( 1.0,5.0 \right)$ &  \\
$\sigma_4$ (km/s) & $U\left( 2.0,8.0 \right)$ &  \\
$f_1$ & $U\left( 0.5,1.0 \right)$ &  \\
$f_2$ & $U\left( 0.0,1.0 \right)$ &  \\
$f_3$ & $U\left( 0.2,1.0 \right)$ &  \\
$f_4$ & $U\left( 0.0,1.0 \right)$ &  \\
$d \lambda_1$ (nm) & $U\left( 3.47,3.481 \right)$ & $\lambda$ shift of K-line in order 7.\\
$d \lambda_2$ (nm) & $U\left( 3.47,3.481 \right)$ & $\lambda$ shift of K-line in order 8. \\
$c_0$ & $U\left( 0.0,2.0 \right)$ & \\
$c_1$ & $U\left( -0.1,0.0 \right)$ & Negative slope. \\
$c_2$ & $U\left( -0.001,0.001 \right)$ &  \\
\end{tabular}
\end{center}
\end{table}

\begin{figure}
    \centering
    \includegraphics[width=0.5 \textwidth,trim={1.7cm 5cm 2.2cm 6cm},clip]{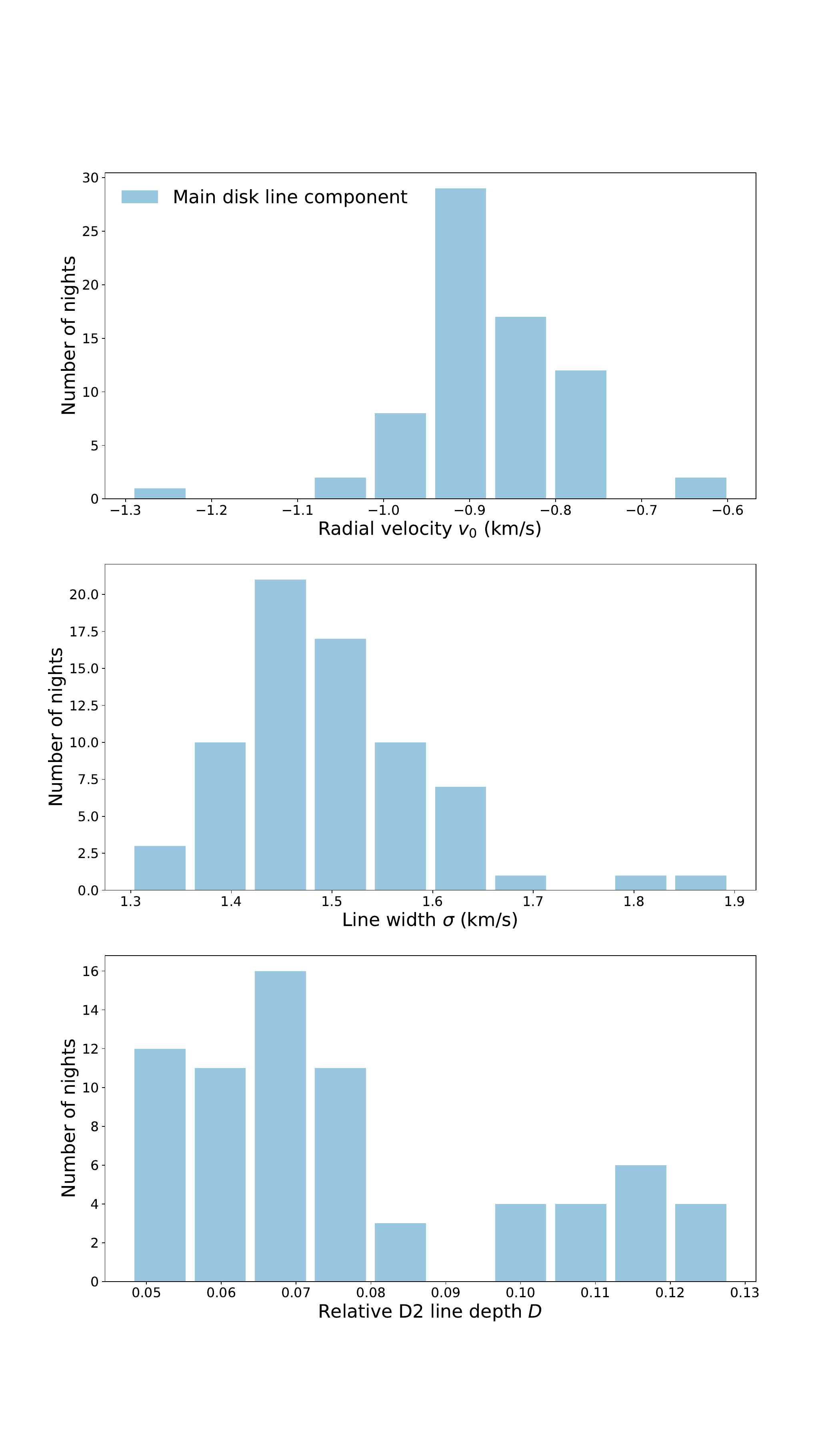}
    \caption{Best-fit radial velocities (top panel), line widths (middle panel) and depths (bottom panel) of the main component of the D2 disk line.}
    \label{fig:Na_disk_distributions}
\end{figure}

\begin{figure}[!htb]
    \centering
    \includegraphics[width=0.5 \textwidth,trim={2cm 1cm 0.5cm 2.5cm},clip]{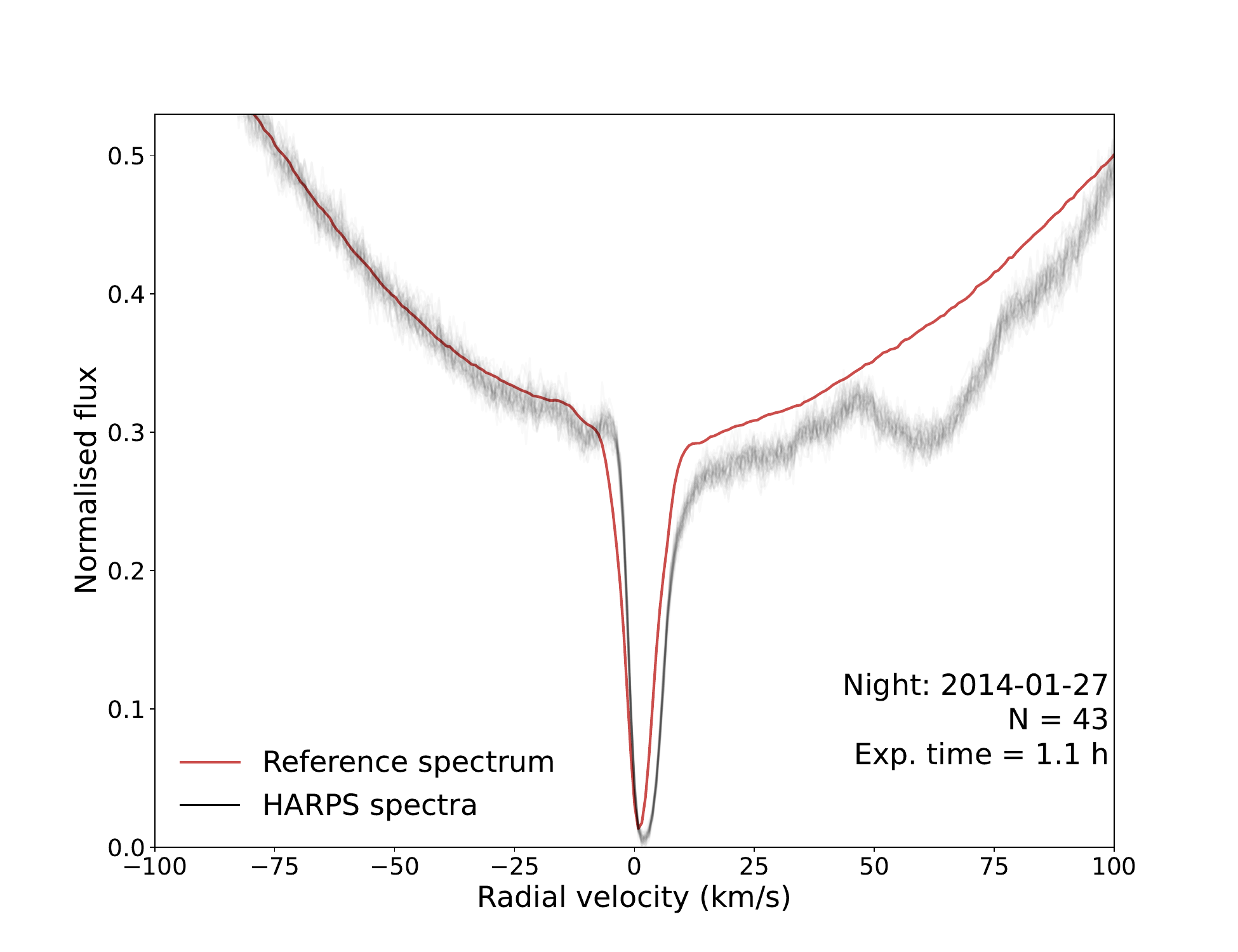}
    \caption{One night of observation in which the Ca\,II disk line is significantly narrower and offset compared to the reference spectrum, indicating intrinsic variability in the disk line.}
    \label{fig:CaII_variability}.
\end{figure}

\begin{figure*}[!htb]
    \centering
    \includegraphics[width=0.48\textwidth,trim={3.5cm 8.2cm 3cm 10.6cm},clip]{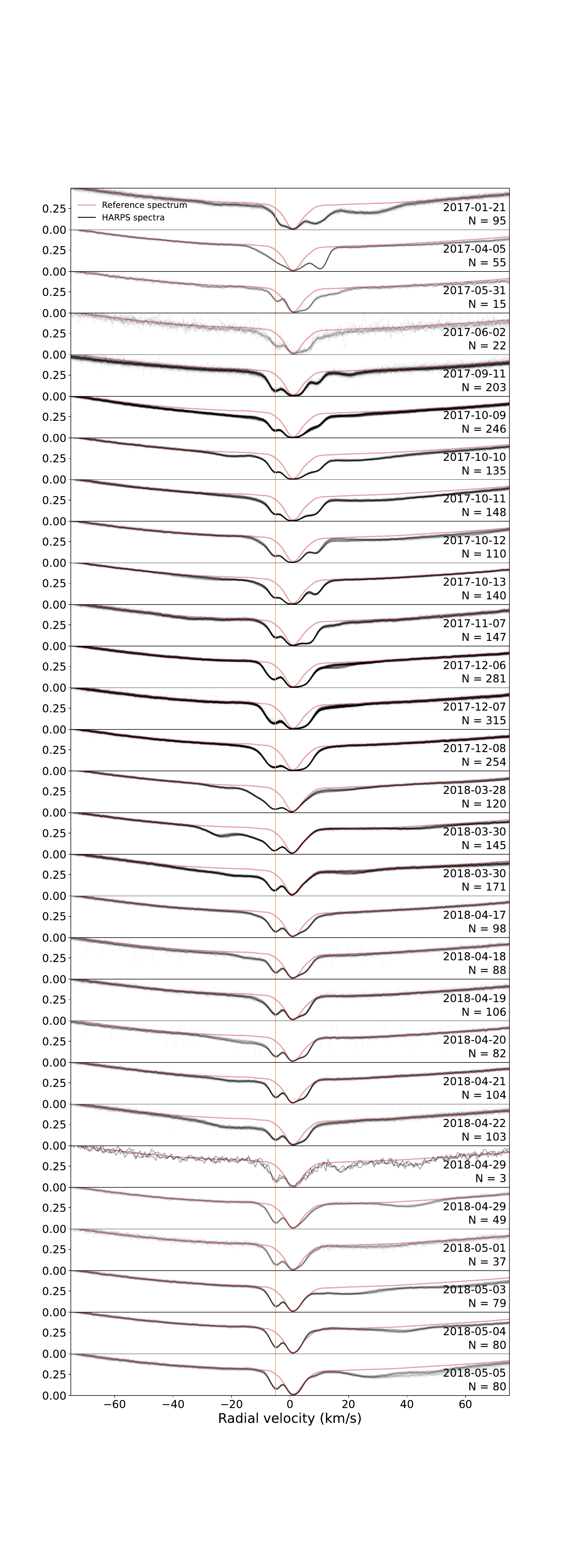}
    \caption{Spectra taken from early 2017 until mid-2018 compared to the reference spectrum. In May of 2017 a blue-shifted feature appeared that lasted for at least a year, indicated with the orange line at -5 km/s.}
    \label{fig:Ca_coherence}
\end{figure*}

\end{appendix}

\end{document}